\documentclass{article} % For LaTeX2e
\usepackage{iclr2017_conference,times}
\usepackage{hyperref}
\usepackage{url}
\usepackage{amsmath,amsfonts,amsthm}
\usepackage{algorithm}% http://ctan.org/pkg/algorithms
\usepackage{algpseudocode}% http://ctan.org/pkg/algorithmicx
\usepackage{dsfont}
\usepackage{graphicx}
\usepackage{bbm}

\title{Maximum Entropy Flow Networks}

\author{Gabriel Loaiza-Ganem\thanks{These authors contributed equally.} , Yuanjun Gao$^\ast$ \& John P. Cunningham \\
Department of Statistics\\
Columbia University\\
New York, NY 10027, USA \\
\texttt{\{gl2480,yg2312,jpc2181\}@columbia.edu}
}

\DeclareMathOperator*{\supp}{supp}

\newcommand{\ms}{\scriptscriptstyle}
\iclrfinalcopy % Uncomment for camera-ready version

\usepackage{mathtools} % pmatrix*! use \begin{matrix*}[r] to center colums right

\newcommand{\optionrule}{\noindent\rule{1.0\textwidth}{0.75pt}}
\renewcommand{\eqref}[1]{eq.~\ref{eq:#1}}

\usepackage{forloop}
\newcommand{\defvec}[1]{\expandafter\newcommand\csname v#1\endcsname{{\mathbf{#1}}}}
\newcounter{ct}
\forLoop{1}{26}{ct}{
    \edef\letter{\alph{ct}}
    \expandafter\defvec\letter
}
\newcommand{\defmat}[1]{\expandafter\newcommand\csname m#1\endcsname{{\mathbf{#1}}}}
\forLoop{1}{26}{ct}{
    \edef\letter{\Alph{ct}}
    \expandafter\defmat\letter
}

  \newtheoremstyle{evandefinition}{\topsep}{\topsep}%
     {}%         Body font  (\itshape)
     {}%         Indent amount (empty = no indent, \parindent = para indent)
     {\bfseries}% Thm head font
     {}%        Punctuation after thm head
     {\newline}%  Space after thm head  (default: 5pt plus 1pt minus 1pt)
     {\thmname{#1}\thmnumber{ #2}. \textit{\thmnote{ #3} }}%         Thm head spec

     \newtheoremstyle{indenteddefinition}{\topsep}{\topsep}
     {\addtolength{\leftskip}{2em}} %         Body font  (\itshape)
     {-1.75em}%         Indent amount (empty = no indent, \parindent = para indent)
     {\bfseries}% Thm head font % also: \scshape
     {}%        Punctuation after thm head
     { }%  Space after thm head  (default: 5pt plus 1pt minus 1pt)
     {\thmname{#1} \thmnumber{#2}. \textbf{(\thmnote{#3})}}%         Thm head spec
\theoremstyle{evandefinition}

\theoremstyle{indenteddefinition}

\theoremstyle{remark}

\makeatletter
\renewcommand*\env@matrix[1][c]{\hskip -\arraycolsep
  \let\@ifnextchar\new@ifnextchar
  \array{*\c@MaxMatrixCols #1}}
\makeatother

\begin{document}

\maketitle

\begin{abstract}
Maximum entropy modeling is a flexible and popular framework for formulating statistical models given partial knowledge. In this paper, rather than the traditional method of optimizing over the continuous density directly, we learn a smooth and invertible transformation that maps a simple distribution to the desired maximum entropy distribution. Doing so is nontrivial in that the objective being maximized (entropy) is a function of the density itself.  By exploiting recent developments in normalizing flow networks, we cast the maximum entropy problem into a finite-dimensional constrained optimization, and solve the problem by combining stochastic optimization with the augmented Lagrangian method. Simulation results demonstrate the effectiveness of our method, and applications to finance and computer vision show the flexibility and accuracy of using maximum entropy flow networks.
\end{abstract}

\section{Introduction}
The maximum entropy (ME) principle \citep{jaynes1957information} states that subject to some given prior knowledge, typically some given list of moment constraints, the distribution that makes minimal additional assumptions -- and is therefore appropriate for a range of applications from hypothesis testing to price forecasting to texture synthesis -- is that which has the largest entropy of any distribution obeying those constraints.  First introduced in statistical mechanics by \cite{jaynes1957information}, and considered both celebrated and controversial, ME has been extensively applied in areas including natural language processing \citep{berger1996maximum}, ecology \citep{phillips2006maximum}, finance \citep{buchen1996maximum}, computer vision \citep{zhu1998filters}, and many more.

Continuous ME modeling problems typically include certain expectation constraints, and are usually solved by introducing Lagrange multipliers, which under typical assumptions yields an exponential family distribution (also called Gibbs distribution) with natural parameters such that the expectation constraints are obeyed.  Unfortunately, fitting ME distributions in even modest dimensions poses significant challenges.    First, optimizing the Lagrangian for a Gibbs distribution requires evaluating the normalizing constant, which is in general computationally very costly and error prone.  Secondly, in all but the rarest cases, there is no way to draw samples independently and identically from this Gibbs distribution, even if one could derive it.  Third, unlike in the discrete case where a number of recent and exciting works have addressed the problem of estimating entropy from discrete-valued data \citep{jiao2015minimax, valiant2013estimating}, estimating differential entropy from data samples remains inefficient and typically biased.  These shortcomings are critical and costly, given the common use of ME distributions for generating reference data samples for a null distribution of a test statistic.  There is thus ample need for a method that can both solve the ME problem and produce a solution that is easy and fast to sample.

In this paper we develop maximum entropy flow networks (MEFN), a stochastic-optimization-based framework and algorithm for fitting continuous maximum entropy models. Two key steps are required.  First, conceptually, we replace the idea of maximizing entropy over a density directly with maximizing, over the parameter space of an indexed function family, the entropy of the density induced by mapping a simple distribution (a Gaussian) through that optimized function.  Modern neural networks, particularly in variational inference  \citep{kingma2013auto, rezende2015variational}, have successfully employed this same idea to generate complex distributions, and we look to similar technologies. Secondly, unlike most other objectives in this network literature, the entropy objective itself requires evaluation of the target density directly, which is unavailable in most traditional architectures.  We overcome this potential issue by learning a smooth, invertible transformation that maps a simple distribution to an (approximate) ME distribution. Recent developments in normalizing flows \citep{rezende2015variational, dinh2016density} allow us to avoid biased and computationally inefficient estimators of differential entropy (such as the nearest-neighbor class of estimators like that of Kozachenko-Leonenko; see \cite{berrett2016efficient}).  Our approach avoids calculation of normalizing constants by learning a map with an easy-to-compute Jacobian, yielding tractable probability density computation. The resulting transformation also allows us to reliably generate iid samples from the learned ME distribution.  We demonstrate MEFN in detail in examples where we can access ground truth, and then we demonstrate further the ability of MEFN networks in equity option prices fitting and texture synthesis. 

Primary contributions of this work include: \emph{(i)} addressing the substantial need for methods to sample ME distributions; \emph{(ii)} introducing ME problems, and the value of including entropy in a range of generative modeling problems, to the deep learning community; \emph{(iii)} the novel use of \emph{constrained} optimization for a deep learning application; and \emph{(iv)} the application of MEFN to option pricing and texture synthesis, where in the latter we show significant increase in the diversity of synthesized textures (over current state of the art) by using MEFN.

%TALK ABOUT MOTIVATION. MAKE SURE TO MENTION WE WILL ONLY CONSIDER THE CONTINUOUS CASE.

\section{Background}
\subsection{Maximum entropy modeling and Gibbs distribution}
We consider a continuous random variable $\mZ \in \mathcal{Z}\subseteq \mathbb{R}^d$ with density $p$, where $p$ has differential entropy $H(p)=- \int p(\vz)\log p(\vz) d\vz$ and support $\supp(p)$. 
%ME problems arise when problem specifics or data dictate a set of known expectations and a support constraint.
%, which here without loss of generality we denote as the constraint $E_{\mZ\sim p}[T(\mZ)] =0$.  
The goal of ME modeling is to find, and then be able to easily sample from, the maximum entropy distribution given a set of moment and support constraints, namely the solution to:
\begin{eqnarray}\label{origprom}
%\substack{E_{\mZ\sim p}[T(\mZ)]=0 \\ \supp(p)=\mathcal{X}}
p^{*} &=& \text{maximize}~~~ H(p) \\
&& \text{subject to} ~~~E_{\mZ\sim p}[T(\mZ)]=0 \nonumber\\
&& ~~~~~~~~~~~~~~~~~~\supp(p)=\mathcal{Z} \nonumber,
\end{eqnarray}

where $T(\vz) = (T_1(\vz),...,T_m(\vz)):\mathcal{Z}\rightarrow \mathbb{R}^{m}$ is the vector of known (assumed sufficient) statistics, and $\mathcal{Z}$ is the given support of the distribution. Under standard regularity conditions, the optimization problem can be solved by Lagrange multipliers, yielding an exponential family $p^*$ of the form:

\begin{equation}\label{curmet}
p^*(\vz) \propto e^{\eta^\top T(\vz)} \mathbbm{1}(\vz \in \mathcal{Z})
\end{equation}

where $\eta \in \mathbb{R}^m$ is the choice of natural parameters of $p^*$ such that $E_{p^*}[T(\mZ)]=0$. 
%Current methods do not reformulate \ref{origprom} as \ref{optobj}, but instead try to find $\eta$ through optimization. 
Despite this simple form, these distributions are only in rare cases tractable from the standpoint of calculating $\eta$, calculating the normalizing constant of $p^*$, and sampling from the resulting distribution. There is extensive literature on finding $\eta$ numerically \citep{darroch1972gicllm, salakhutdinov2003conopt, dellapietra1997ifrf, dudik2004maxentdenest, malouf2002maxentcomp,collins2002lrabbd}, but doing so requires computing normalizing constants, which poses a challenge even for problems with modest dimensions. Also, even if $\eta$ is correctly found, it is still not trivial to sample from $p^*$. Problem-specific sampling methods (such as importance sampling, MCMC, etc.) have to be designed and used, which is in general challenging (burn-in, mixing time, etc.) and computationally burdensome.  

%and MCMC methods have to be used. This requires problem-specific considerations and faces the usual MCMC issues: having to wait a burn-in period, the chains might mix very slowly so the target distribution might not be sampled from, and having to throw away samples because they are correlated instead of iid. Reformulating \ref{origprom} as \ref{optobj} solves this issue. Once $f^*$ is found, it is extremely easy to get an iid sample from $p^*$: just get an iid sample from $p_{\ms 0}$ and run it through $f^*$.

\subsection{Normalizing flows}
Following \cite{rezende2015variational}, we define a \textit{normalizing flow} as the transformation of a probability density through a sequence of invertible mappings. Normalizing flows provide an elegant way of generating a complicated distribution while maintaining tractable density evaluation. Starting with a simple distribution $\mZ_0 \in \mathbb{R}^d \sim p_0$ (usually taken to be a standard multivariate Gaussian), and by applying $k$ invertible and smooth functions $f_i: \mathbb{R}^d \rightarrow \mathbb{R}^d (i=1,...,k)$, the resulting variable $\mZ_{k} = f_{k}\circ f_{k-1}\circ\dots \circ f_{1}(\mZ_{0})$ has density:
\begin{equation}
p_{k}(\vz_{k}) = p_{\ms 0}(f_{1}^{-1}\circ f_{2}^{-1}\circ\dots\circ f_{k}^{-1}(\vz_{k}))\displaystyle \prod_{i=1}^{k}|\det(J_{i}(\vz_{i-1}))|^{-1},
\end{equation}
where $J_i$ is the Jacobian of $f_i$. If the determinant of $J_{i}$ can be easily computed, $p_{k}$ can be computed efficiently. 

\cite{rezende2015variational} proposed two specific families of transformations for variational inference, namely planar flows and radial flows, respectively:
\begin{equation}
f_i(\vz) = \vz + \vu_i h(\vw_i^{T} \vz+ b_i)~~~~~~~~~~\text{and}~~~~~~~~~~f_i(\vz) = \vz + \beta_i h(\alpha_i, r_i)(\vz - \vz'_i),
\end{equation}
where $b_i \in \mathbb{R}$, $\vu_i, \vw_i \in \mathbb{R}^d$ and $h$ is an activation function in the planar case, and 
%. Radial flows are functions of the form:
%\begin{equation}
%f_i(\vz) = \vz + \beta_i h(\alpha_i, r_i)(\vz - \vz'_i),
%\end{equation}
where $\beta_i \in \mathbb{R}$, $\alpha_i >0$, $\vz'_i \in \mathbb{R}^d$ , $h(\alpha,r)=1/(\alpha+r)$ and $r_i = ||\vz-\vz'_i||$ in the radial. %For a certain domain of the parameters, the transformations are invertible and the determinant of the Jacobian can be easily computed with the matrix determinant lemma. %\cite{rezende2015variational} illustrates that incorporating normalizing flows in variational inference framework improves the performance.
Recently \cite{dinh2016density} proposed a normalizing flow with convolutional, multiscale structure that is suitable for image modeling and has shown promise in density estimation for natural images. 

\section{Maximum entropy flow network (MEFN) algorithm}
\subsection{Formulation}
Instead of solving Equation \ref{curmet}, we propose solving Equation \ref{origprom} directly by optimizing a transformation that maps a random variable $\mZ_0$, with simple distribution $p_0$, to the ME distribution. Given a parametric family of normalizing flows $\mathcal{F}=\{f_\phi, \phi \in \mathbb{R}^q\}$, we denote $p_\phi(\vz) = p_0( f_\phi^{-1}(\vz) ) | \det (J_\phi (\vz)) |^{-1}$ as the distribution of the variable $f_\phi(\mZ_0)$, where $J_\phi$ is the Jacobian of $f_\phi$. We then rewrite the ME problem as:
%and $t\in \mathbb{R}^{n}$. To do so, we will sample $Z_{\ms 0}$ from a distribution $p_{\ms 0}$ which is easy to sample from and transform $Z_{\ms 0}$ by applying a function $f$ so that the resulting distribution is $p^{*}$. We thus reformulate the problem in the following way:
%\begin{equation}\label{optobj}
%\phi^{*} = \argmax_{\substack{E[T(f_\phi(\mZ_{0}))]=0 \\ \supp(p_\phi) \in \mathcal{X}}} H(p_\phi)
%\end{equation}
\begin{eqnarray}\label{optobj}
%\substack{E_{\mZ\sim p}[T(\mZ)]=0 \\ \supp(p)=\mathcal{X}}
\phi^{*} &=& \text{maximize}~~~ H(p_\phi) \\
&& \text{subject to} ~~~E_{\mZ_0\sim p_0}[T(f_\phi(\mZ_{0}))]=0 \nonumber\\
&& ~~~~~~~~~~~~~~~~~~\supp(p_\phi)=\mathcal{Z}. \nonumber
\end{eqnarray}

When $p_0$ is continuous and $\mathcal{F}$ is suitably general, the program in Equation \ref{optobj} recovers the ME distribution $p_\phi$ exactly. With a flexible transformation family, the ME distribution can be well approximated. 
%In our applications we take $\mathcal{F}$ to be a composition of planar flows ($10$ layers) followed by an invertible function $g$ whose image is $\mathcal{Z}$. The choice of $g$ should be done in such a way that the determinant of its Jacobian is tractable. For most distributions of interest $g$ is not difficult to find and allows the support to be removed from the learning process.  
In experiments we found that taking $p_0$ to be a standard multivariate normal distribution achieves good empirical performance. Taking $p_0$ to be a bounded distribution (e.g. uniform distribution) is problematic for learning transformations near the boundary, and heavy tailed distributions (e.g. Cauchy distribution) caused similar trouble due to large numbers of outliers. %For all the experiments we take $p_0$ to be a multivariate standard normal distribution.

\subsection{Algorithm}
\label{sec:c_update}

We solved Equation \ref{optobj} using the augmented Lagrangian method. Denote $R(\phi) = E(T(f_\phi( \mZ_0)))$, the augmented Lagrangian method uses the following objective:

\begin{equation}\label{auglag}
L(\phi; \lambda, c) = - H(p_\phi) + \lambda^{\top} R(\phi) + \dfrac{c}{2}||R(\phi)||^2
\end{equation}

where $\lambda \in \mathbb{R}^m$ is the Lagrange multiplier and $c>0$ is the penalty coefficient. We minimize Equation $\ref{auglag}$ for a non-decreasing sequence of $c$ and well-chosen $\lambda$. As a technical note, the augmented Lagrangian method is guaranteed to converge under some regularity conditions \citep{bertsekas1996}. As is usual in neural networks, a proof of these conditions is challenging and not yet available, though intuitive arguments (see Appendix \S \ref{sec:augLcond}) suggest that most of them should hold.  Due to the non rigorous nature of these arguments, we rely on the empirical results of the algorithm to claim that it is indeed solving the optimization problem.

For a fixed $(\lambda, c)$ pair, we optimize $L$ with stochastic gradient descent. Owing to our choice of network and the resulting ability to efficiently calculate the density $p_\phi(\vz^{(i)})$ for any sample point $\vz^{(i)}$ (which are easy-to-sample iid draws from the multivariate normal $p_0$), we compute the unbiased estimator of $H(p_\phi)$ with:
\begin{equation}
H(p_\phi) \approx -\frac{1}{n} \sum_{i=1}^n \log p_\phi ( f_\phi(\vz^{(i)}) )
\end{equation}

$R(\phi)$ can also be estimated without bias by taking a sample average of $\vz^{(i)}$ draws.  The resulting optimization procedure is detailed in Algorithm \ref{alg:maxent}, of which step 9 requires some detail: denoting $\phi_k$ as the resulting $\phi$ after $i_{max}$ SGD iterations at the augmented Lagrangian iteration $k$, the usual update rule for $c$ \citep{bertsekas1996} is:
\begin{equation}\label{cupdates}
c_{k+1}=\begin{cases}
\beta c_{k}\text{, if }||R(\phi_{k+1})||>\gamma ||R(\phi_{k})||\\
c_{k}\text{, otherwise}
\end{cases}
\end{equation}
where $\gamma\in (0,1)$ and $\beta>1$. Monte Carlo estimation of $R(\phi)$ sometimes caused $c$ to be updated too fast, causing numerical issues. Accordingly, we changed the hard update rule for $c$ to a probabilistic update rule: a hypothesis test is carried out with null hypothesis $H_{0} : E[||R(\phi_{k+1})||]=E[\gamma||R(\phi_{k})||]$ and alternative hypothesis $H_{1} : E[||R(\phi_{k+1})||]>E[\gamma||R(\phi_{k})||]$. The $p$-value $p$ is computed, and $c_{k+1}$ is updated to $\beta c_{k}$ with probability $1-p$. We used a two-sample $t$-test to calculate the $p$-value.
%
%In our experiments, we found $n=300$, $\tilde{n}=1000$, $\beta=4$, $\gamma=0.25$, $k_{max}=10$ and $i_{max}=3000$ to give good performance.
What results is a robust and novel algorithm for estimating maximum entropy distributions, while preserving the critical properties of being both easy to calculate densities of particular points, and being trivially able to produce truly iid samples.
%To solve the constrained optimization problem we propose using a stochastic version of augmented Lagrangian multiplier method 

%This way, $f^*(Z_{\ms 0}) \sim p^{*}$. Since optimizing over all functions can't be done analytically for most choices of $\mathcal{X}$ and $T$, we will optimize over a function class $\mathcal{F}$ parametrized by $\phi$. As is conventional in the optimization literature, we will consider instead the minimization problem:

%\begin{equation}\label{optobjmin}
%f^{*} = \argmin_{\substack{E[T(f(Z_{\ms 0}))]-t=0 \\ f(Z_{\ms 0})\in \mathcal{X}}} -H(f(Z_{\ms 0}))
%\end{equation}

\begin{algorithm}[t]
\caption{Training the MEFN}
\begin{algorithmic}[1]
\State initialize $\phi = \phi_0$, set $c_0 > 0$ and $\lambda_0$.
\For{Augmented Lagrangian iteration $k = 1,...,k_{\max}$} 
\For{SGD iteration $i = 1,...,i_{\max}$}
\State{Sample $\vz^{(1)},...,\vz^{(n)} \sim p_0$, get transformed variables $\vz^{(i)}_\phi = f_\phi(\vz^{(i)}), i = 1,...,n$}
\State{Update $\phi$ by descending its stochastic gradient (using e.g. ADADELTA \citep{zeiler2012adadelta}):
{\scriptsize \[\nabla_\phi L(\phi; \lambda_k, c_k) \approx   \frac{1}{n} \sum_{i=1}^n \nabla_\phi  \log p_\phi ( \vz^{(i)}_\phi  ) + 
 \frac{1}{n} \sum_{i=1}^n \nabla_\phi T( \vz^{(i)}_\phi  ) \lambda_k +  c_k  \frac{2}{n} \sum_{i=1}^{\frac{n}{2}} \nabla_\phi T( \vz^{(i)}_\phi  ) \cdot \frac{2}{n}\sum_{i=\frac{n}{2}+1}^n T(  \vz^{(i)}_\phi   )  \]}}
\EndFor
\State Sample $\vz^{(1)},...,\vz^{(\tilde{n})} \sim p_0$, get transformed variables $\vz^{(i)}_\phi = f_\phi(\vz^{(i)}), i = 1,...,\tilde{n}$
\State Update $\lambda_{k+1} = \lambda_k + c_k \frac{1}{\tilde{n}} \sum_{i=1}^{\tilde{n}} T( \vz^{(i)}_\phi )$
\State Update $c_{k+1}\geq c_{k}$ (see text for detail)
\EndFor
%\Until{\textbf{convergence}}
\end{algorithmic}
\label{alg:maxent}
\end{algorithm}

%\begin{algorithm}[t]
%\label{alg:maxens}
%\caption{Maximum Entropy Sampling Networks}
%\begin{algorithmic}[1]
%\FOR{<condition>} \STATE {<text>} \ENDFOR

%\State Initialize $i=0$, $\phi_0$, $c_0 > 0$ and $\lambda_0$
%\Repeat
%\For{i training iterations}
%\State Find $\phi_{i+1} = \argmin L_{c_i}(\phi,\lambda_i)$ using stochastic optimization and $\phi_{i}$ as the initial point
%\State Update $\lambda_{i+1} = \lambda_i + c_i T_M(\phi_{i+1})$ where $M\geq m$.
%\State Take $c_{i+1} \geq c_{i}$
%\State $i \leftarrow i+1$
%\Until{\textbf{convergence}}
%\end{algorithmic}
%\end{algorithm}

\section{Experiments}

We first construct an ME problem with a known solution (\S \ref{sec:simulation}), and we analyze the MEFN algorithm with respect to the ground truth and to an approximate Gibbs solution.  These examples test the validity of our algorithm and illustrate its performance. \S \ref{sec:real} and \S \ref{sec:texture} then applies the MEFN to a financial data application (predicting equity option values) and texture synthesis, respectively, to illustrate the flexibility and practicality of our algorithm.

For \S \ref{sec:simulation} and \S \ref{sec:real}, We use 10 layers of planar flow with a final transformation $g$ (specified below) that transforms samples to the specified support, and use with ADADELTA \citep{zeiler2012adadelta}. For \S \ref{sec:texture} we use real NVP structure and use ADAM \citep{kingma2014adam} with learning rate = $0.001$. For all our experiments, we use $i_{max}=3000$, $\beta=4$, $\gamma=0.25$. For \S \ref{sec:simulation} and \S \ref{sec:real} we use $n=300$, $\tilde{n}=1000$, $k_{\max}=10$; For \S \ref{sec:texture} we use $n=\tilde{n}=2$, $k_{\max}=8$. % to give good performance

 \subsection{A maximum entropy problem with known solution}
 \label{sec:simulation}
 
Following the setup of the typical ME problem, suppose we are given a specified support $\mathcal{S}=\{\vz = (z_1,\dots,z_{d-1}):z_i\geq 0\text{ and }\sum_{k=1}^{d-1}z_k\leq 1\}$ and a set of constraints $E[\log Z_k] = \kappa_k (k = 1,...,d)$, where $Z_d = 1 - \sum_{k=1}^{d-1} Z_k$.  We then write the maximum entropy program:
\begin{eqnarray}\label{dirichletprob}
%\substack{E_{\mZ\sim p}[T(\mZ)]=0 \\ \supp(p)=\mathcal{X}}
p^{*} &=& \text{maximize}~~~ H(p) \\
&& \text{subject to} ~~~E_{\mZ\sim p}[\log Z_k - \kappa_k]=0 ~~~ \forall k=1,...,d\nonumber\\
&& ~~~~~~~~~~~~~~~~~~\supp(p)=\mathcal{S}\nonumber.%\left\{(z_1,\dots,z_{d-1}):z_k\geq 0\text{ and }\sum_{k=1}^{d-1}z_k\leq 1\right\} \nonumber.
\end{eqnarray}
 
This is a general ME problem that can be solved via the MEFN.  Of course, we have particularly chosen this example because, though it may not obviously appear so, the solution has a standard and tractable form, namely the Dirichlet.  This choice allows us to consider a complicated optimization program that happens to have known global optimum, providing a solid test bed for the MEFN (and for the Gibbs approach against which we will compare).   Specifically, given a parameter $\alpha\in\mathbb{R}^d$, the Dirichlet has density:
%To verify that our method works, we test it to sample from Dirichlet and Wishart distributions. 
\begin{equation}
p(z_1,\dots,z_{d-1})=\dfrac{1}{B(\alpha)}\prod_{k=1}^{d}z_k^{\alpha_k-1}\mathbbm{1}\left( (z_1,\dots,z_{d-1}) \in \mathcal{S} \right)
\end{equation}
where $B(\alpha)$ is the multivariate Beta function, and $z_d=1-\sum_{k=1}^{d-1}z_k$. Note that this Dirichlet is a distribution on $\mathcal{S}$ and not on the $(d-1)$-dimensional simplex $\mathcal{S}^{d-1}=\{(z_1,\dots,z_d):z_k\geq 0\text{ and }\sum_{k=1}^{d}z_k= 1\}$ (an often ignored and seemingly unimportant technicality that needs to be correct here to ensure the proper transformation of measure). 
%Although seemingly an unimportant technicality, this will actually be relevant. 
%
Connecting this familiar distribution to the ME problem above, we simply have to choose $\alpha$ such that $\kappa_k = \psi(\alpha_k)-\psi(\alpha_0)$ for $k=1,...,d$, where $\alpha_0=\sum_{k=1}^{d}\alpha_k$ and $\psi$ is the digamma function.  We then can pose the above ME problem to the MEFN and compare performance against ground truth.
Before doing so, we must stipulate the transformation $g$ that maps the Euclidean space of the multivariate normal $p_0$ to the desired support $\mathcal{S}$.  Any sensible choice will work well (another point of flexibility for the MEFN); we use the standard transformation:
\begin{equation}
 g(z_1,...,z_{d-1}) = 
\left(\dfrac{e^{z_1}}{\sum_{k=1}^{d-1}e^{z_k}+1}, ..., \dfrac{e^{z_{d-1}}}{\sum_{k=1}^{d-1}e^{z_k}+1} \right)^\top
\end{equation}
%
%It is the maximum entropy distribution on $S$ subject to the constraints:
%\begin{equation}\label{dircon}
%E[\log(Z_i)]=\psi(\alpha_i)-\psi(\alpha_0)\text{ for }i=1,\dots,d
%\end{equation}
%where $\psi$ is the digamma function and $\alpha_0=\sum_{i=1}^{n}\alpha_i$. We take $g$ to be:
%
%\begin{equation}
%g\left(
%\begin{array}{c}
%z_1\\
%\vdots\\
%z_{d-1}\\
%\end{array}
%\right)=\left(
%\begin{array}{c}
%\dfrac{e^{z_1}}{\sum_{k=1}^{d-1}e^{z_k}+1}\\
%\vdots\\
%\dfrac{e^{z_{d-1}}}{\sum_{k=1}^{d-1}e^{z_k}+1}\\
%\end{array}
%\right)
%\end{equation}
Note that the MEFN outputs vectors in $\mathbb{R}^{d-1}$, and not $\mathbb{R}^d$, because the Dirichlet is specified as a distribution on $\mathcal{S}$ (and not on the simplex $\mathcal{S}^{d-1}$). Accordingly, the Jacobian is a square matrix and its determinant can be computed efficiently using the matrix determinant lemma.
%(if the output was in $\mathbb{R}^d$ we would need to compute the Gram matrix \cite{benisrael1999chvarmat} instead of the Jacobian). 
Here, $p_0$ is set to the $(d-1)$-dimensional standard normal.\\

We proceed as follows: We choose $\alpha$ and compute the constraints $\kappa_1,...,\kappa_{d}$.  We run MEFN pretending we do not know $\alpha$ or the Dirichlet form. We then take a random sample from the fitted distribution and a random sample from the Dirichlet with parameter $\alpha$, and compare the two samples using the maximum mean discrepancy (MMD) kernel two sample test \cite[]{gretton2012kerneltest}, which assesses the fit quality.  We take the sample size to be $300$ for the two sample kernel test. Figure 1 shows an example of the transformation from normal (left panel) to MEFN (middle panel), and comparing that to the ground truth Dirichlet (right panel).  The MEFN and ground truth Dirichlet densities shown in purple match closely, and the samples drawn (red) indeed appear to be iid draws from the same (maximum entropy) distribution in both cases. 

Additionally, the middle panel of Figure 1 shows an important cautionary tale that foreshadows our texture synthesis results (\S \ref{sec:texture}).  One might suppose that satisfying the moment matching constraints is adequate to produce a distribution which, if not technically the ME distribution, is still interestingly variable.  The middle panel shows the failure of this intuition: in dark green, we show a network trained to simply match the moments specified above, and the resulting distribution quite poorly expresses the variability available to a distribution with these constraints, leading to samples that are needlessly similar.  Given the substantial interest in using networks to learn implicit generative models (e.g., \cite{mohamed2016learning}), this concern is particularly relevant and highlights the importance of considering entropy.

\begin{figure}[htbp]
\label{fig:dir}
\centering
\begin{tabular}[t]{ccc}
\vspace{-0cm}
Initial distribution $p_0$& MEFN result $p_{\phi^*}$&Ground truth $p^*$\\
\includegraphics[scale=0.4,clip = true, trim=1cm 1.5cm 1cm 1.5cm]{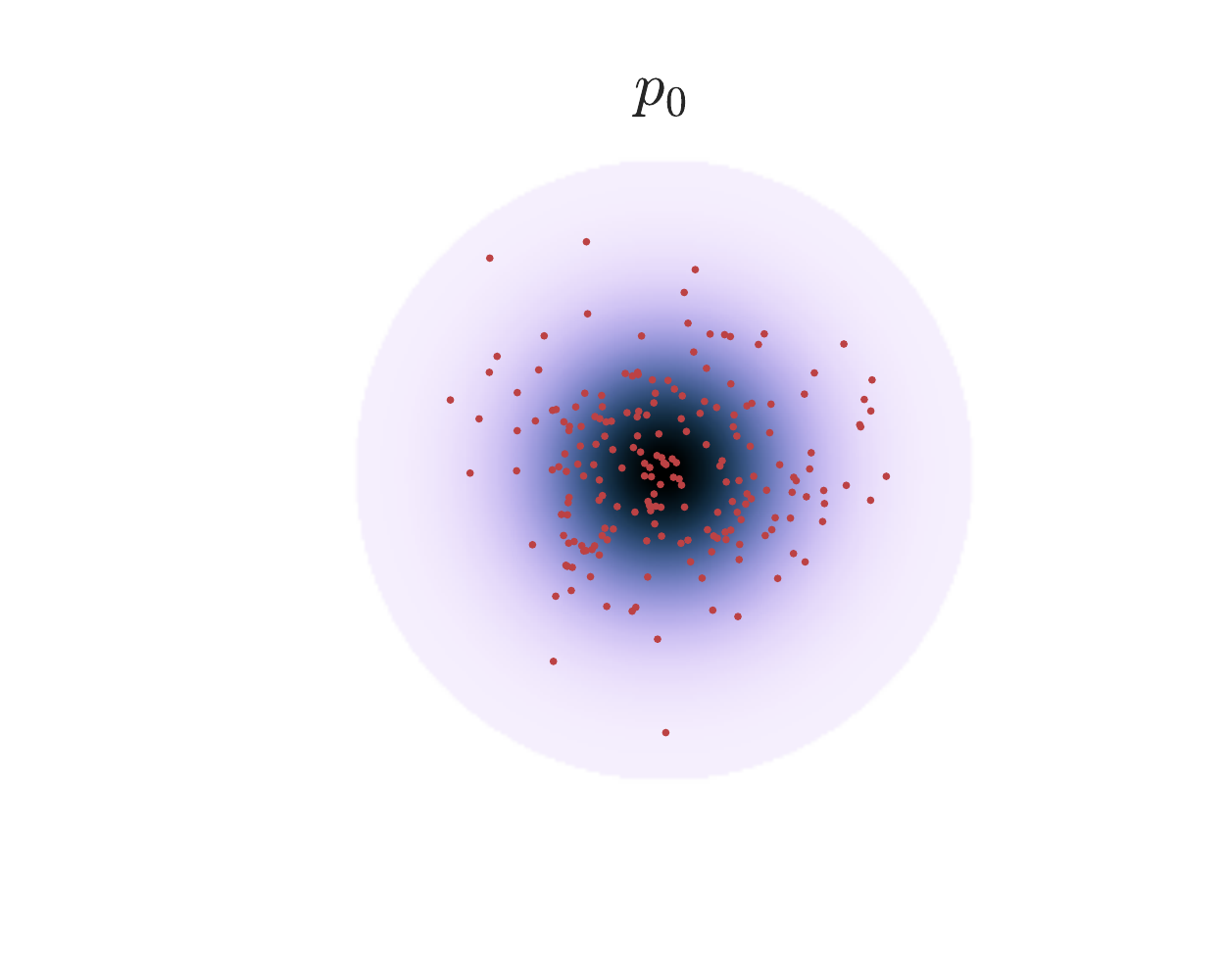}&
\includegraphics[scale=0.4,clip = true, trim=1cm 1.5cm 1cm 1.5cm]{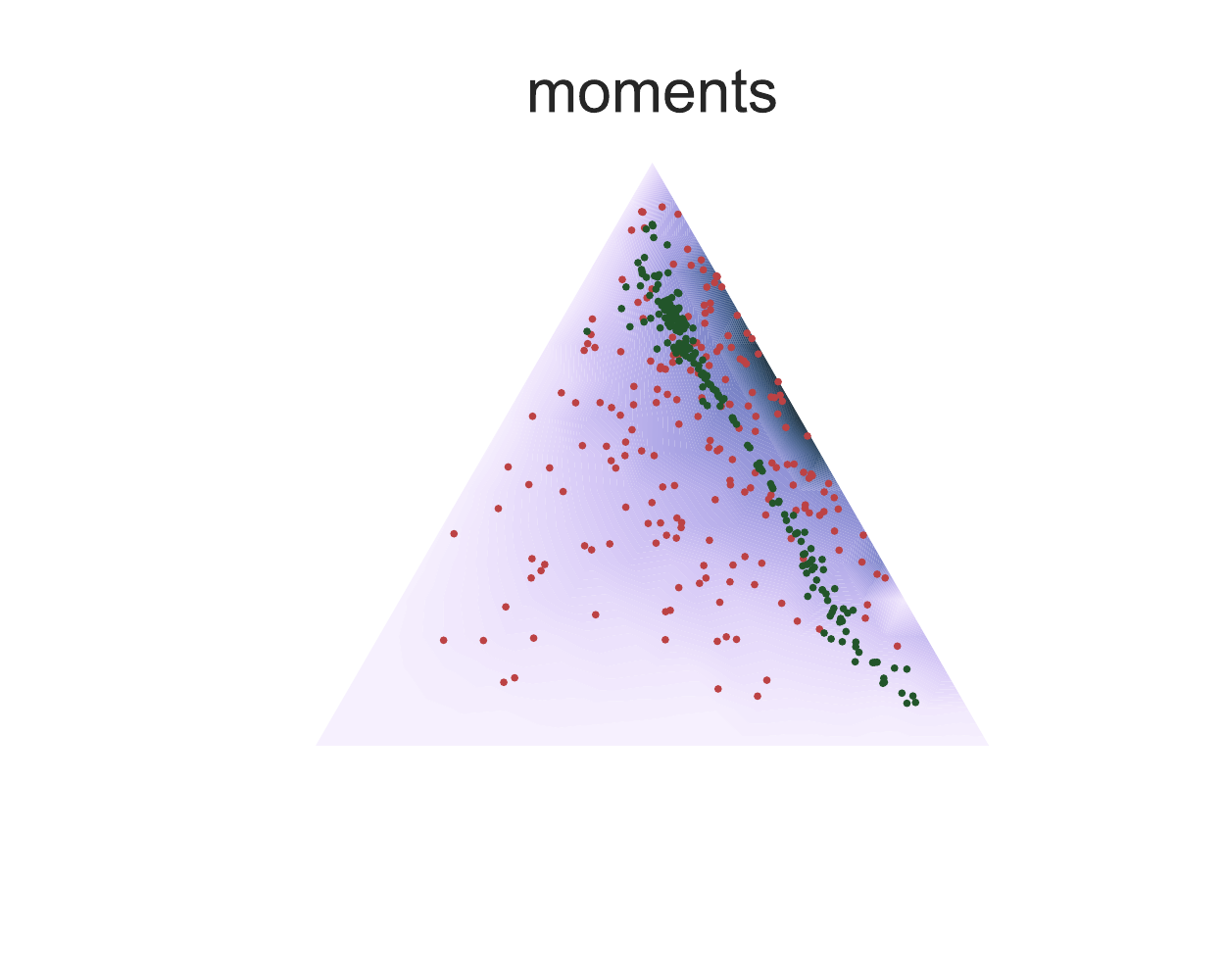}&
\includegraphics[scale=0.4,clip = true, trim=1cm 1.5cm 1cm 1.5cm]{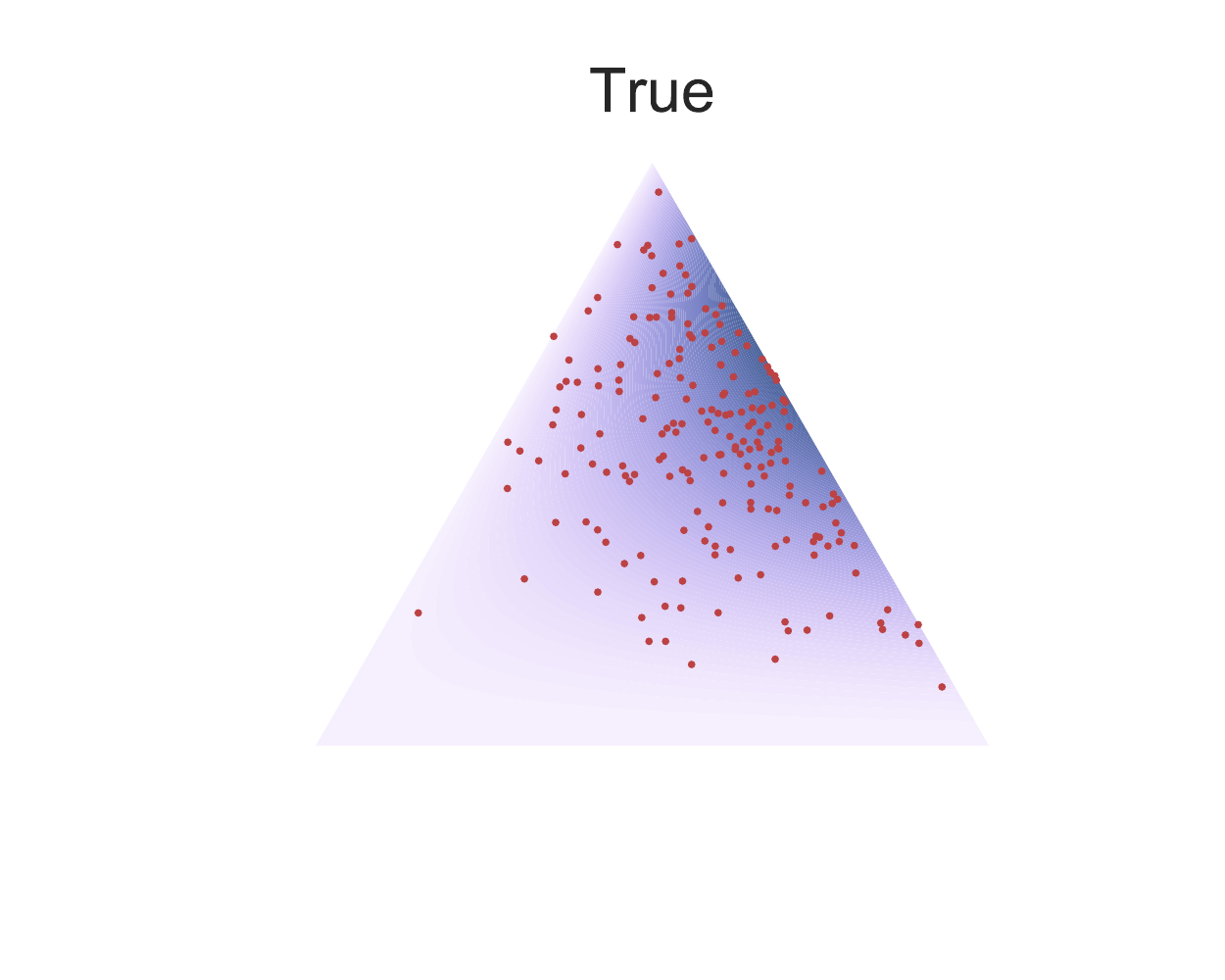}\\
%\\
\end{tabular}
\caption{Example results from the ME problem with known Dirichlet ground truth. \textit{Left panel}: The normal density $p_0$ (purple) and iid samples from $p_0$ (red points).  \textit{Middle panel}: The MEFN transforms $p_0$ to the desired maximum entropy distribution $p_{\phi^*}$ on the simplex (calculated density $p_{\phi^*}$ in purple).  Truly iid samples are easily drawn from $p_{\phi^*}$ (red points) by drawing from $p_0$ and mapping those points through $f_{\phi^*}$.  Shown in the middle panel are the same points in the top left panel mapped through $f_{\phi^*}$. Samples corresponding to training the same network as MEFN to simply match the specified moments (ignoring entropy) are also shown (dark green points; see text). \textit{Right panel}: The ground truth (in this example, known to be Dirichlet) distribution in purple, and iid samples from it in red.}
\end{figure}

\begin{figure}[!ht]
\label{fig:dir2}
\centering
\begin{tabular}[t]{ccc}
\vspace{-0cm}
\includegraphics[scale=0.35,clip = true]{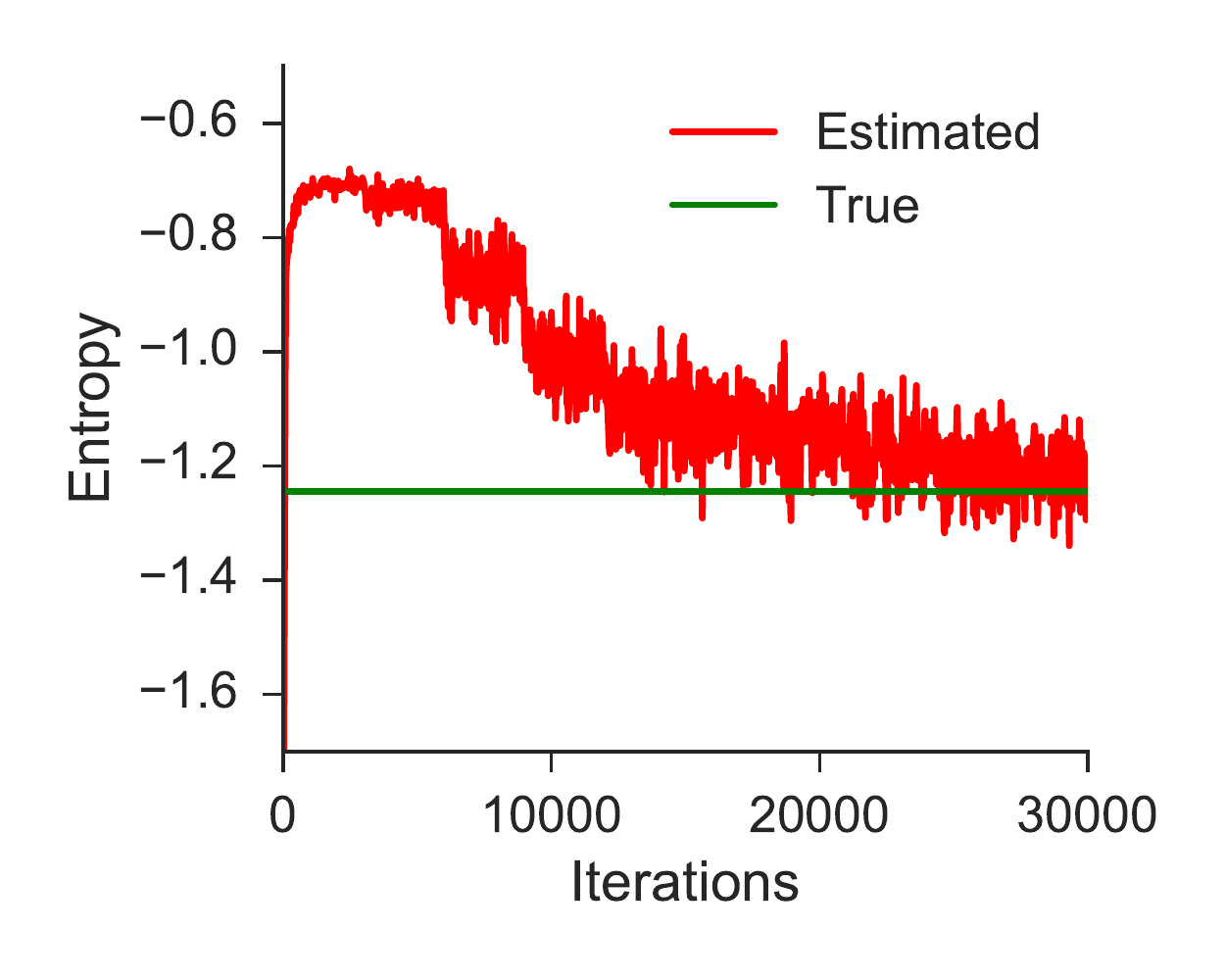}&
\includegraphics[scale=0.35,clip = true, trim=0cm 0cm 0cm 0cm]{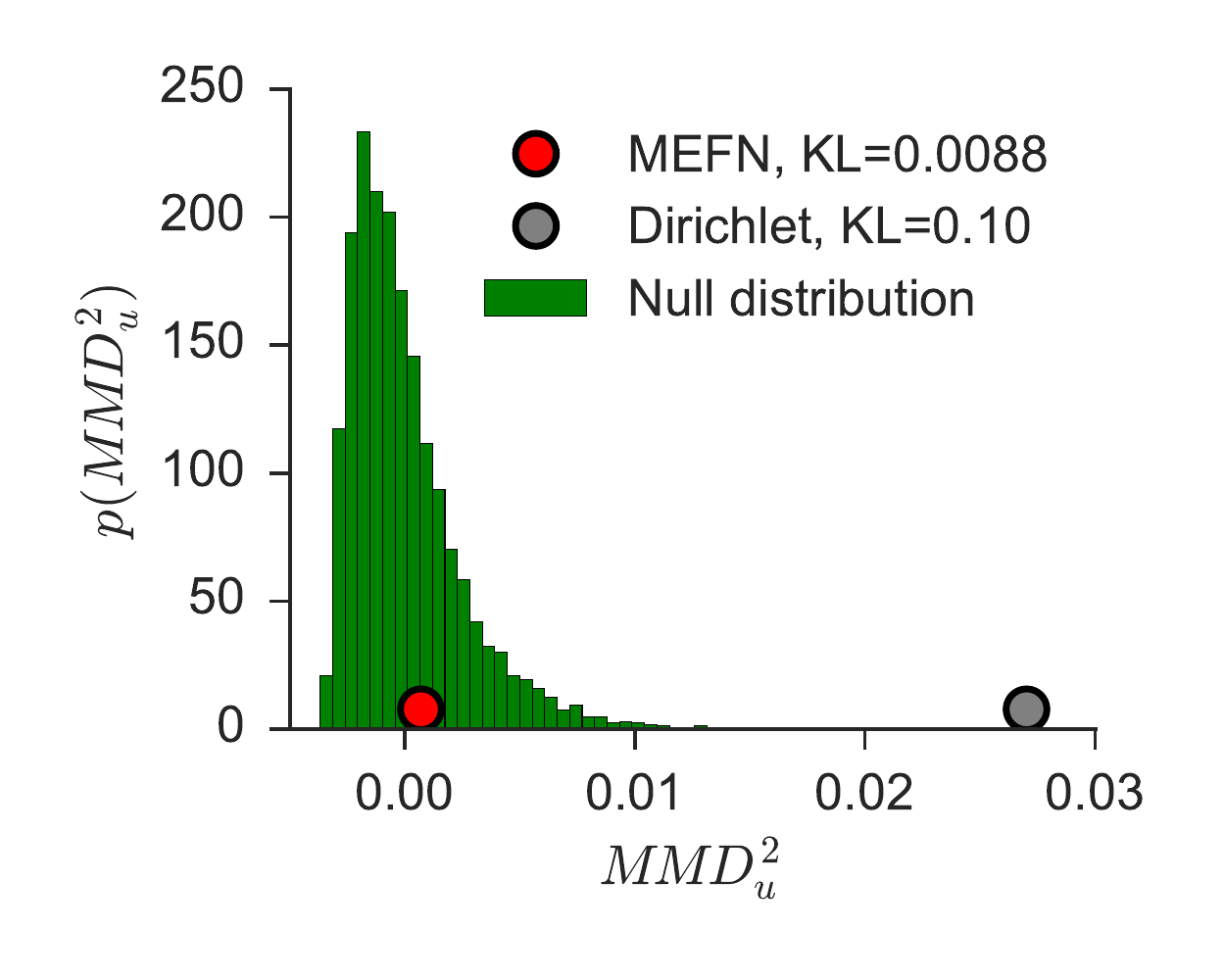}&
\includegraphics[scale=0.35,clip = true, trim=0cm 0cm 0cm 0cm]{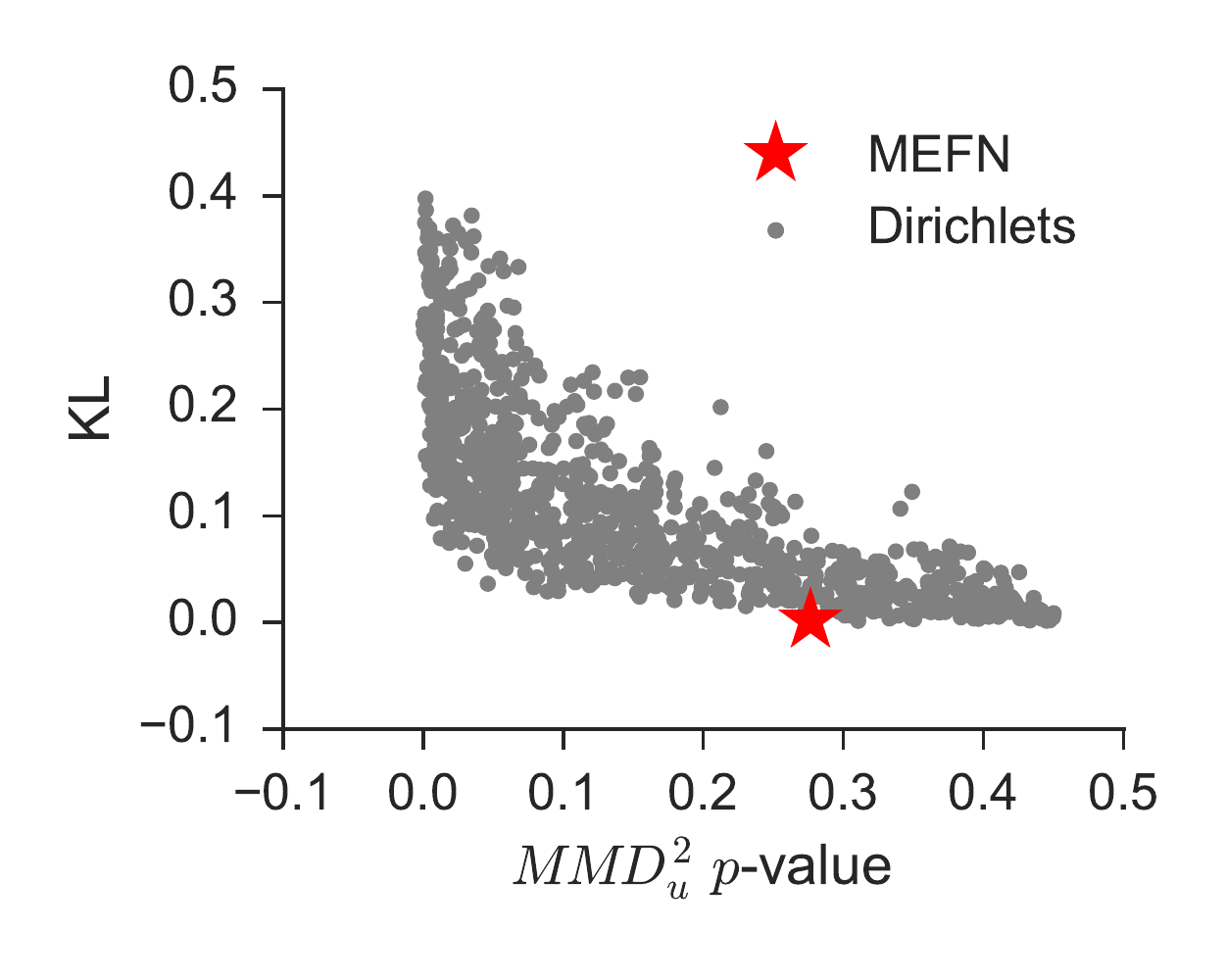}\\
\end{tabular}
\vspace{-0.5cm}
\caption{Quantitative analysis of simulation results. See text for description.}
\end{figure}

Figure 2 quantitatively analyzes these results.  In the left panel, for a specific choice of $\alpha=(1,2,3)$, we show our unbiased entropy estimate of the MEFN distribution $p_\phi$ as a function of the number of SGD iterations (red), along with the ground truth maximum entropy $H(p^*)$ (green line).  Note that the MEFN stabilizes at the correct value (as a stochastic estimator, variance around that value is expected).   In the middle panel, we show the distribution of MMD values for the kernel two sample test, as well as the observed statistic for the MEFN (red) and for a randomly chosen Dirichlet distribution (gray; chosen to be close to the true optimum, making a conservative comparison).  The MMD test does not reject MEFN as being different from the true ME distribution $p^*$, but it does reject a Dirichlet whose $KL$ to the true $p^*$ is small (see legend). In the right panel, for many different Dirichlets in a small grid around a single true $p^*$, the kernel two sample test statistic is computed, the MMD $p$-value is calculated, as is  the $KL$ to the true distribution. We plot a scatter of these points in grey, and we plot the particular MEFN solution as a red star. We see that for other Dirichlets with similar $KL$ to the true distribution as the MEFN distribution, the $p$-values seem uniform, meaning that the $KL$ to the true is indeed very small.  Again this is conservative, as the grey points have access to the known Dirichlet form, whereas the MEFN considered the entire space (within its network capacity) of $\mathcal{S}$ supported distributions.  Given this fact, the performance of MEFN is impressive.

%In the above example We tried different distributions of $\mZ_{0}$ in our experiments and found the best performance was achieved when using a standard multivariate Gaussian of the appropriate size. For example, if each coordinate of $\mZ_{0}$ was uniform, our method would have more difficulties learning where it should send the coordinate when it is near 0 or 1. If each coordinate of $\mZ_{0}$ was Cauchy, outliers would be present far too often and would not be mapped correctly. The normal distribution is nice because it does not have issues at the boundaries like the uniform and it has light tails, unlike the Cauchy. For this reason, we take $\mZ_{ 0}$ to be standard multivariate normal in what follows.

\subsection{Risk-neutral asset pricing}
\label{sec:real}
We illustrate the flexibility and practicality of our algorithm extracting the risk-neutral asset price probability based on option prices, an active and interesting area for ME models. We find that MEFN and the classic Gibbs approach yield comparable performances. Owing to space limitations we have placed these results in Appendix \S \ref{sec:real}.

\subsection{Modeling images of textures}
\label{sec:texture}
Constructing generative models to generate random images with certain texture structure is an important task in computer vision. A line of texture synthesis research proceeds by first extracting a set of features that characterizes the target texture and then generate images that match the features. The seminal work of \cite{zhu1998filters} proposes constructing texture models under the ME framework, where features (or filters) of the given texture image are adaptively added in the model and a Gibbs distribution whose expected feature matches the target texture is learnt. One major difficulty with the method is that both model learning and image generation involve sampling from a complicated Gibbs distribution. 
%, and while MCMC methods are proposed, the performance can be unsatisfying for large images and complicated features. 
More recent works exploit more complicated features \citep{portilla2000parametric, gatys2015texture, ulyanov2016texture}. \cite{ulyanov2016texture} propose the \emph{texture net}, which uses a texture loss function by using the Gram matrices of the outputs of some convolutional layers of a pre-trained deep neural network for object recognition. 

While the use of these complicated features does provide high-quality synthetic texture images, that work focuses exclusively on generating images that match these feature (moments).  Importantly, this network focuses only on generating feature-matching images without using the ME framework to promote the diversity of the samples.  Doing so can be deeply problematic: in Figure 1 (middle panel), we showed the lack of diversity resulting from only moment matching in that Dirichlet setting, and further we note that the extreme pathology would result in a point mass on the training image -- a global optimum for this objective, but obviously a terrible generative model for synthesizing textures.  Ideally, the MEFN will match the moments \emph{and} promote sample diversity. %Indeed, both evaluating the entropy and sampling from ME distribution are challenging in the high-dimensional image space.

We applied MEFN to texture synthesis with an RGB representation of the $224 \times 224$ pixel images , $\vz \in \mathcal{Z} = [0,1]^{d}$, where $d=224\times 224 \times 3$. We follow \cite{ulyanov2016texture} (we adapted \url{https://github.com/ProofByConstruction/texture-networks}) to create a texture loss measure $T(\vz): [0,1]^{d} \rightarrow \mathbb{R}$, 
%\footnote{We adapted the publicly available TensorFlow implementation from \url{https://github.com/ProofByConstruction/texture-networks}; see \cite{ulyanov2016texture}.}
 and aim to sample a diverse set of images with small moment violation. For the transformation family $\mathcal{F}$ we use the real NVP network structure proposed in \cite{dinh2016density} (we adapted \url{https://github.com/taesung89/real-nvp}). We use $3$ residual blocks with $32$ feature maps for each coupling layer and downscale $3$ times.
 %\footnote{We adapted \url{https://github.com/taesung89/real-nvp}; see \cite{dinh2016density}.}
%We compare our MEFN formulation (Algorithm \ref{alg:maxent}) with the formulation of \cite{ulyanov2016texture}, $\hat{\phi} = \arg \min_{\phi} E_{\mZ_0\sim p_0} [ T(f_\phi(\mZ_0)) ]$ (Equation (7) of \cite{ulyanov2016texture}), which corresponds to the second term of Equation \ref{auglag} and only aims at minimizing the expected texture loss. 
For fair comparison, we use the same real NVP structure for both\footnote{\cite{ulyanov2016texture} use a quite different generative network structure, which is not invertible and is therefore infeasible for entropy evaluation, so we replace their generative network by the real NVP structure.}, implemented in TensorFlow \citep{abadi2016tensorflow}.

\begin{figure}[!ht]
\centering
\begin{tabular}[t]{ccc}
Input & Texture net \citep{ulyanov2016texture} & MEFN (ours) \\
%\vspace{-0cm}
\includegraphics[scale=0.43,clip = true]{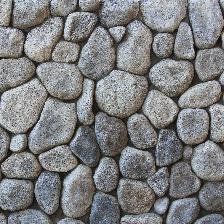}&
\includegraphics[scale=0.43,clip = true]{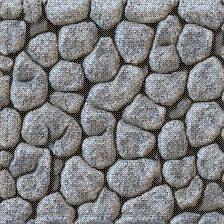}&
\includegraphics[scale=0.43,clip = true, trim=0cm 0cm 0cm 0cm]{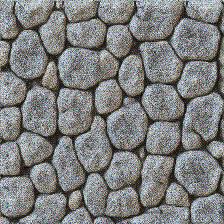}\\
Texture cost & Entropy & RGB histogram\\
\includegraphics[scale=0.43,clip = true]{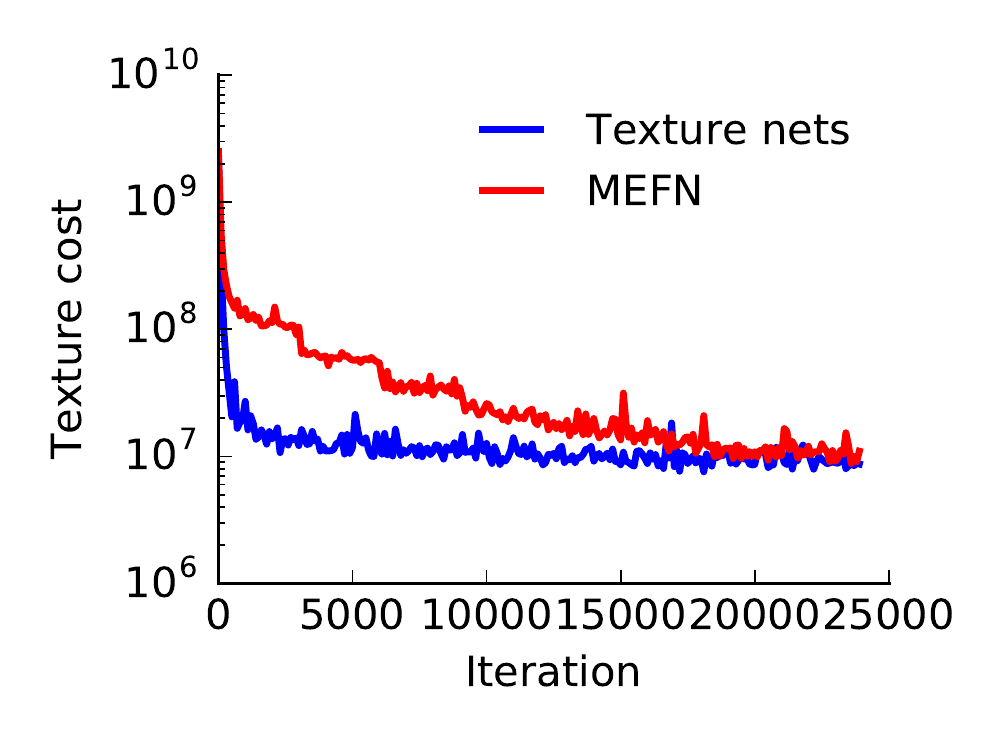}&
\includegraphics[scale=0.43,clip = true]{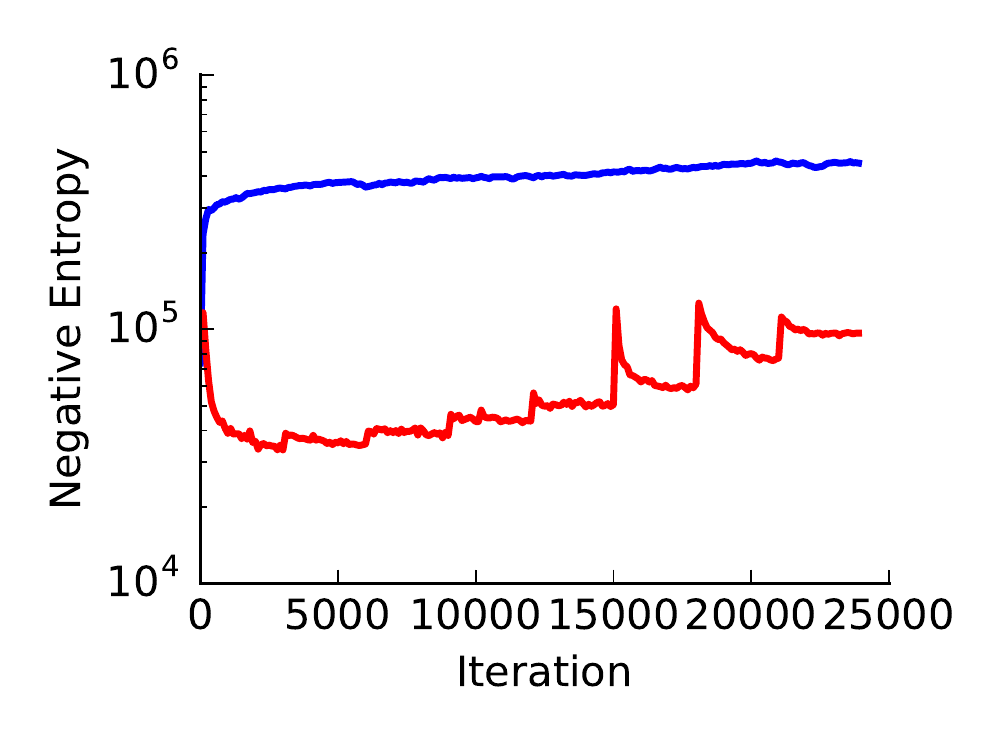}&
\includegraphics[scale=0.43,clip = true]{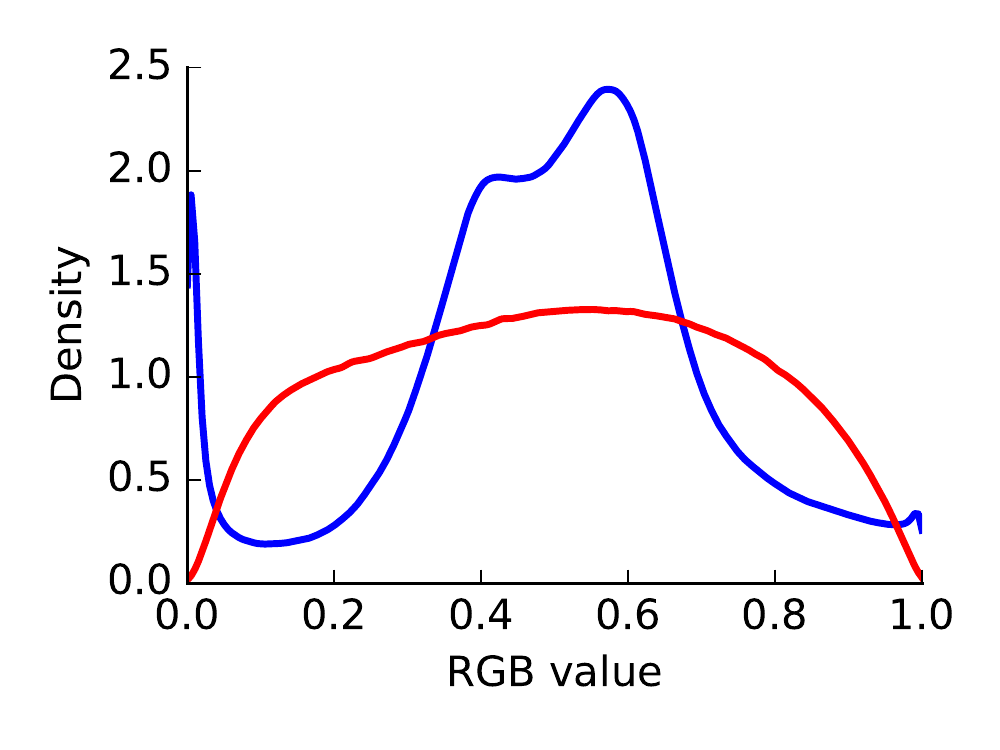} \\
%\\
\end{tabular}
\vspace{-0.5cm}
\caption{Analysis of texture synthesis experiment. See text for description.}
\label{fig:texture}
%\label{fig:reaching_projection}
\end{figure}

As is shown in top row of figure \ref{fig:texture}, both methods generate visually pleasing images capturing the texture structure well. The bottom row of Figure \ref{fig:texture} shows that texture cost (left panel) is similar for both methods, while MEFN generates figures with much larger entropy than the texture network formulation (middle panel), which is desirable (as previously discussed). The bottom right panel of figure \ref{fig:texture} compares the marginal distribution of the RGB values sampled from the networks: we found that MEFN generates a more variable distribution of RGB values than the texture net. Further results are in Appendix \S \ref{sec:textapp}.

We compute in Table \ref{tab:texture} the average pairwise Euclidean distance between randomly sampled images ($d_{L^2} = \text{mean}_{i \neq j} \|\vz_i - \vz_j \|_2^2$), and MEFN gives higher $d_{L^2}$, quantifying diversity across images.
%For $20$ randomly sampled images, we get $d_{L^2} = 11534$ for texture net and $d_{L^2} = 17014$ for MEFN. 
We also consider an ANOVA-style analysis to measure the diversity of the images, where we think of the RGB values for the same pixel across multiple images as a group, and compute the within and between group variance. Specifically, denoting $z_i^k$ as the pixel value for a specific pixel $k = 1,...,d$ for an image $i = 1,....,n$. We partition the total sum of square $\text{SST} = \sum_{i,k} (z_i^k - \bar{z})^2$ as the within group error $\text{SSW} = \sum_{i,k} (z_i^k - \bar{z}^k)^2$ and between group error $\text{SSB} = \sum_{i,k} n (\bar{z}^k - \bar{z})^2$, where $\bar{z}$ and $\bar{z}^k$ are the mean pixel values across all data and for a specific pixel $k$. Ideally we want the samples to exhibit large variability across images (large SSW, within a group/pixel) and no structure in the mean image (small SSB, across groups/pixels).  Indeed, the MEFN has a larger SSW, implying higher variability around the mean image,
 a smaller SSB, implying the stationarity of the generated samples, and a larger SST, implying larger total variability also.  The MEFN produces images that are conclusively more variable without sacrificing the quality of the texture, implicating the broad utility of ME.
%Maybe we can change this from table to text. Though maybe not much space would be saved (YG)
\begin{table}[htpb]
\caption{Quantitative measure of image diversity using $20$ randomly sampled images}
\label{sample-table}
\begin{center}
\begin{tabular}{l | l | lll}
Method  & $d_{L^2}$& SST & SSW & SSB \\
\hline 
Texture net  & 11534 & 128680 & 109577 & 19103\\
MEFN   & 17014 & 175604 & 161639 & 13964 \\
\end{tabular}
\end{center}
\label{tab:texture}
\end{table}
\section{Conclusion}
%In this paper we propose a general framework for fitting ME models. By learning a transformation of a simple distribution rather than the distribution itself, we are able to get around computing the normalizing constant for the distribution. By combining stochastic optimization with the augmented Lagrangian method, we can fit the model efficiently. The resulting distribution approximates the ME distribution well, and we can easily evaluate the density and draw iid samples from the distribution. We illustrate in both a simulated case with known ground truth and a real data example case that our method works well and successfully obtains the ME solution.
%Future work for the algorithm involves identifying more expressive transformation families (e.g. \cite{kingma2016improving}) to further increase the flexibility and accuracy of the model. While a finite number of layers may not exactly extract the ME distribution, the learned resulting distribution has some regularization effect that can strike a balance between accuracy and smoothness, which can be desirable in cases when either the constraints are noisy or a smooth distribution is desirable. 
%In terms of applications, we are excited to explore structured high-dimensional ME models in time series and spatial modeling (e.g. \cite{zhu1998filters}). We would expect structured transformations incorporating the temporal or spatial homogeneity to be efficient and expressive in fitting ME models in these domains.
In this paper we propose a general framework for fitting ME models. This approach is novel and has three key features. First, by learning a transformation of a simple distribution rather than the distribution itself, we are able to avoid explicitly computing an intractable normalizing constant for the ME distribution. Second, by combining stochastic optimization with the augmented Lagrangian method, we can fit the model efficiently, allowing us to evaluate the ME density of any point simply and accurately. Third, critically, this construction allows us to trivially sample iid from a ME distribution, extending the utility and efficiency of the ME framework more generally. Also, accuracy equivalent to the classic Gibbs approach is in itself a contribution (owing to these other features). We illustrate the MEFN in both a simulated case with known ground truth and real data examples. %case, where our method works well and successfully obtains the ME solution.

There are a few recent works encouraging sample diversity in the setting of texture modeling. \cite{ulyanov2017improved} extended \cite{ulyanov2016texture} by adding a penalty term using the Kozachenko-Leonenko estimator \cite{kozachenko1987sample} of entropy. Their generative network is an arbitrary deep neural network rather than a normalizing flow, which is more flexible but cannot give the probability density of each sample easily so as to compute an unbiased estimator of the entropy. Kozachenko-Leonenko is a biased estimator for entropy and requires a fairly large number of samples to get good performance in high-dimensional settings, hindering the scalability and accuracy of the method; indeed, our choice of normalizing flow networks was driven by these practical issues with Kozachenko-Leonenko. \cite{LuZhuWu2016} extended \cite{zhu1998filters} by using a more flexible set of filters derived from a pre-trained deep neural networks, and using parallel MCMC chains to learn and sample from the Gibbs distribution. Running parallel MCMC chains results in diverse samples but can be computationally intensive for generating each new sample image. Our MEFN framework enables truly iid sampling with the ease of a feed forward network.

%Future work for the algorithm involves identifying more expressive transformation families (e.g.\cite{kingma2016improving}) to further increase the flexibility and accuracy of the model, while maintaining the key Jacobian properties we exploited here. While a finite number of layers may not exactly extract the ME distribution, the learned resulting distribution has some regularization effect that can strike a balance between accuracy and smoothness, which can be desirable in cases when either the constraints are noisy or a smooth distribution is desirable. %In terms of applications, we seek to explore structured high-dimensional ME models in time series and spatial modeling (e.g.  \cite{zhu1998filters}). We would expect structured transformations incorporating temporal or spatial homogeneity to be efficient and expressive in fitting ME models in these domains.

\subsubsection*{Acknowledgments}

We thank Evan Archer for normalizing flow code, and Xuexin Wei, Christian Andersson Naesseth and Scott Linderman for helpful discussion. This work was supported by a Sloan Fellowship and a McKnight Fellowship (JPC).

\newpage

\bibliography{../bibliography}

\begin{thebibliography}{36}
\providecommand{\natexlab}[1]{#1}
\providecommand{\url}[1]{\texttt{#1}}
\expandafter\ifx\csname urlstyle\endcsname\relax
  \providecommand{\doi}[1]{doi: #1}\else
  \providecommand{\doi}{doi: \begingroup \urlstyle{rm}\Url}\fi

\bibitem[Abadi et~al.(2016)Abadi, Agarwal, Barham, Brevdo, Chen, Citro,
  Corrado, Davis, Dean, Devin, et~al.]{abadi2016tensorflow}
Mart{\i}n Abadi, Ashish Agarwal, Paul Barham, Eugene Brevdo, Zhifeng Chen,
  Craig Citro, Greg~S Corrado, Andy Davis, Jeffrey Dean, Matthieu Devin, et~al.
\newblock Tensorflow: Large-scale machine learning on heterogeneous distributed
  systems.
\newblock \emph{arXiv preprint arXiv:1603.04467}, 2016.

\bibitem[Berger et~al.(1996)Berger, Pietra, and Pietra]{berger1996maximum}
Adam~L Berger, Vincent J~Della Pietra, and Stephen A~Della Pietra.
\newblock A maximum entropy approach to natural language processing.
\newblock \emph{Computational linguistics}, 22\penalty0 (1):\penalty0 39--71,
  1996.

\bibitem[Berrett et~al.(2016)Berrett, Samworth, and Yuan]{berrett2016efficient}
Thomas~B Berrett, Richard~J Samworth, and Ming Yuan.
\newblock Efficient multivariate entropy estimation via $ k $-nearest neighbour
  distances.
\newblock \emph{arXiv preprint arXiv:1606.00304}, 2016.

\bibitem[Bertsekas(2014)]{bertsekas1996}
Dimitri~P Bertsekas.
\newblock \emph{Constrained optimization and Lagrange multiplier methods}.
\newblock Academic press, 2014.

\bibitem[Bondarenko(2003)]{bondarenko2003estimation}
Oleg Bondarenko.
\newblock Estimation of risk-neutral densities using positive convolution
  approximation.
\newblock \emph{Journal of Econometrics}, 116\penalty0 (1):\penalty0 85--112,
  2003.

\bibitem[Borwein et~al.(2003)Borwein, Choksi, and
  Mar{\'e}chal]{borwein2003probability}
Jonathan Borwein, Rustum Choksi, and Pierre Mar{\'e}chal.
\newblock Probability distributions of assets inferred from option prices via
  the principle of maximum entropy.
\newblock \emph{SIAM Journal on Optimization}, 14\penalty0 (2):\penalty0
  464--478, 2003.

\bibitem[Buchen \& Kelly(1996)Buchen and Kelly]{buchen1996maximum}
Peter~W Buchen and Michael Kelly.
\newblock The maximum entropy distribution of an asset inferred from option
  prices.
\newblock \emph{Journal of Financial and Quantitative Analysis}, 31\penalty0
  (01):\penalty0 143--159, 1996.

\bibitem[Choromanska et~al.(2015)Choromanska, Henaff, Mathieu, Arous, and
  LeCun]{choromanska2015loss}
Anna Choromanska, Mikael Henaff, Michael Mathieu, G{\'e}rard~Ben Arous, and
  Yann LeCun.
\newblock The loss surfaces of multilayer networks.
\newblock In \emph{AISTATS}, 2015.

\bibitem[Collins et~al.(2002)Collins, Schapire, and Singer]{collins2002lrabbd}
Michael Collins, Robert~E Schapire, and Yoram Singer.
\newblock Logistic regression, adaboost and bregman distances.
\newblock \emph{Machine Learning}, 48\penalty0 (1-3):\penalty0 253--285, 2002.

\bibitem[Darroch \& Ratcliff(1972)Darroch and Ratcliff]{darroch1972gicllm}
John~N Darroch and Douglas Ratcliff.
\newblock Generalized iterative scaling for log-linear models.
\newblock \emph{The annals of mathematical statistics}, pp.\  1470--1480, 1972.

\bibitem[Della~Pietra et~al.(1997)Della~Pietra, Della~Pietra, and
  Lafferty]{dellapietra1997ifrf}
Stephen Della~Pietra, Vincent Della~Pietra, and John Lafferty.
\newblock Inducing features of random fields.
\newblock \emph{IEEE transactions on pattern analysis and machine
  intelligence}, 19\penalty0 (4):\penalty0 380--393, 1997.

\bibitem[Dinh et~al.(2016)Dinh, Sohl-Dickstein, and Bengio]{dinh2016density}
Laurent Dinh, Jascha Sohl-Dickstein, and Samy Bengio.
\newblock Density estimation using real nvp.
\newblock \emph{arXiv preprint arXiv:1605.08803}, 2016.

\bibitem[Dudik et~al.(2004)Dudik, Phillips, and
  Schapire]{dudik2004maxentdenest}
Miroslav Dudik, Steven~J Phillips, and Robert~E Schapire.
\newblock Performance guarantees for regularized maximum entropy density
  estimation.
\newblock In \emph{International Conference on Computational Learning Theory},
  pp.\  472--486. Springer, 2004.

\bibitem[Figlewski(2008)]{figlewski2008estimating}
Stephen Figlewski.
\newblock Estimating the implied risk neutral density.
\newblock 2008.

\bibitem[Gatys et~al.(2015)Gatys, Ecker, and Bethge]{gatys2015texture}
Leon Gatys, Alexander~S Ecker, and Matthias Bethge.
\newblock Texture synthesis using convolutional neural networks.
\newblock In \emph{Advances in Neural Information Processing Systems}, pp.\
  262--270, 2015.

\bibitem[Gretton et~al.(2012)Gretton, Borgwardt, Rasch, Sch{\"o}lkopf, and
  Smola]{gretton2012kerneltest}
Arthur Gretton, Karsten~M Borgwardt, Malte~J Rasch, Bernhard Sch{\"o}lkopf, and
  Alexander Smola.
\newblock A kernel two-sample test.
\newblock \emph{Journal of Machine Learning Research}, 13\penalty0
  (Mar):\penalty0 723--773, 2012.

\bibitem[Jaynes(1957)]{jaynes1957information}
Edwin~T Jaynes.
\newblock Information theory and statistical mechanics.
\newblock \emph{Physical review}, 106\penalty0 (4):\penalty0 620, 1957.

\bibitem[Jiao et~al.(2015)Jiao, Venkat, Han, and Weissman]{jiao2015minimax}
Jiantao Jiao, Kartik Venkat, Yanjun Han, and Tsachy Weissman.
\newblock Minimax estimation of functionals of discrete distributions.
\newblock \emph{IEEE Transactions on Information Theory}, 61\penalty0
  (5):\penalty0 2835--2885, 2015.

\bibitem[Kawaguchi(2016)]{kawaguchi2016deep}
Kenji Kawaguchi.
\newblock Deep learning without poor local minima.
\newblock In \emph{Advances In Neural Information Processing Systems}, pp.\
  586--594, 2016.

\bibitem[Kingma \& Ba(2014)Kingma and Ba]{kingma2014adam}
Diederik Kingma and Jimmy Ba.
\newblock Adam: A method for stochastic optimization.
\newblock \emph{arXiv preprint arXiv:1412.6980}, 2014.

\bibitem[Kingma \& Welling(2013)Kingma and Welling]{kingma2013auto}
Diederik~P Kingma and Max Welling.
\newblock Auto-encoding variational bayes.
\newblock \emph{arXiv preprint arXiv:1312.6114}, 2013.

\bibitem[Kozachenko \& Leonenko(1987)Kozachenko and
  Leonenko]{kozachenko1987sample}
LF~Kozachenko and Nikolai~N Leonenko.
\newblock Sample estimate of the entropy of a random vector.
\newblock \emph{Problemy Peredachi Informatsii}, 23\penalty0 (2):\penalty0
  9--16, 1987.

\bibitem[Lu et~al.(2016)Lu, Zhu, and Wu]{LuZhuWu2016}
Yang Lu, Song-chun Zhu, and Ying~Nian Wu.
\newblock Learning frame models using cnn filters.
\newblock In \emph{Thirtieth AAAI Conference on Artificial Intelligence}, 2016.

\bibitem[Malouf(2002)]{malouf2002maxentcomp}
Robert Malouf.
\newblock A comparison of algorithms for maximum entropy parameter estimation.
\newblock In \emph{proceedings of the 6th conference on Natural language
  learning-Volume 20}, pp.\  1--7. Association for Computational Linguistics,
  2002.

\bibitem[Mohamed \& Lakshminarayanan(2016)Mohamed and
  Lakshminarayanan]{mohamed2016learning}
Shakir Mohamed and Balaji Lakshminarayanan.
\newblock Learning in implicit generative models.
\newblock \emph{arXiv preprint arXiv:1610.03483}, 2016.

\bibitem[Phillips et~al.(2006)Phillips, Anderson, and
  Schapire]{phillips2006maximum}
Steven~J Phillips, Robert~P Anderson, and Robert~E Schapire.
\newblock Maximum entropy modeling of species geographic distributions.
\newblock \emph{Ecological modelling}, 190\penalty0 (3):\penalty0 231--259,
  2006.

\bibitem[Poole et~al.(2016)Poole, Lahiri, Raghu, Sohl-Dickstein, and
  Ganguli]{poole2016exponential}
Ben Poole, Subhaneil Lahiri, Maithreyi Raghu, Jascha Sohl-Dickstein, and Surya
  Ganguli.
\newblock Exponential expressivity in deep neural networks through transient
  chaos.
\newblock In \emph{Advances In Neural Information Processing Systems}, pp.\
  3360--3368, 2016.

\bibitem[Portilla \& Simoncelli(2000)Portilla and
  Simoncelli]{portilla2000parametric}
Javier Portilla and Eero~P Simoncelli.
\newblock A parametric texture model based on joint statistics of complex
  wavelet coefficients.
\newblock \emph{International journal of computer vision}, 40\penalty0
  (1):\penalty0 49--70, 2000.

\bibitem[Raghu et~al.(2016)Raghu, Poole, Kleinberg, Ganguli, and
  Sohl-Dickstein]{raghu2016expressive}
Maithra Raghu, Ben Poole, Jon Kleinberg, Surya Ganguli, and Jascha
  Sohl-Dickstein.
\newblock On the expressive power of deep neural networks.
\newblock \emph{arXiv preprint arXiv:1606.05336}, 2016.

\bibitem[Rezende \& Mohamed(2015)Rezende and Mohamed]{rezende2015variational}
Danilo~Jimenez Rezende and Shakir Mohamed.
\newblock Variational inference with normalizing flows.
\newblock \emph{arXiv preprint arXiv:1505.05770}, 2015.

\bibitem[Salakhutdinov et~al.(2002)Salakhutdinov, Roweis, and
  Ghahramani]{salakhutdinov2003conopt}
Ruslan Salakhutdinov, Sam Roweis, and Zoubin Ghahramani.
\newblock On the convergence of bound optimization algorithms.
\newblock In \emph{Proceedings of the Nineteenth conference on Uncertainty in
  Artificial Intelligence}, pp.\  509--516. Morgan Kaufmann Publishers Inc.,
  2002.

\bibitem[Ulyanov et~al.(2016)Ulyanov, Lebedev, Vedaldi, and
  Lempitsky]{ulyanov2016texture}
Dmitry Ulyanov, Vadim Lebedev, Andrea Vedaldi, and Victor Lempitsky.
\newblock Texture networks: Feed-forward synthesis of textures and stylized
  images.
\newblock \emph{arXiv preprint arXiv:1603.03417}, 2016.

\bibitem[Ulyanov et~al.(2017)Ulyanov, Vedaldi, and
  Lempitsky]{ulyanov2017improved}
Dmitry Ulyanov, Andrea Vedaldi, and Victor Lempitsky.
\newblock Improved texture networks: Maximizing quality and diversity in
  feed-forward stylization and texture synthesis.
\newblock \emph{arXiv preprint arXiv:1701.02096}, 2017.

\bibitem[Valiant \& Valiant(2013)Valiant and Valiant]{valiant2013estimating}
Paul Valiant and Gregory Valiant.
\newblock Estimating the unseen: improved estimators for entropy and other
  properties.
\newblock In \emph{Advances in Neural Information Processing Systems}, pp.\
  2157--2165, 2013.

\bibitem[Zeiler(2012)]{zeiler2012adadelta}
Matthew~D Zeiler.
\newblock Adadelta: an adaptive learning rate method.
\newblock \emph{arXiv preprint arXiv:1212.5701}, 2012.

\bibitem[Zhu et~al.(1998)Zhu, Wu, and Mumford]{zhu1998filters}
Song~Chun Zhu, Yingnian Wu, and David Mumford.
\newblock Filters, random fields and maximum entropy (frame): Towards a unified
  theory for texture modeling.
\newblock \emph{International Journal of Computer Vision}, 27\penalty0
  (2):\penalty0 107--126, 1998.

\end{thebibliography}
\bibliographystyle{iclr2017_conference}

\newpage

\appendix

\section{Augmented Lagrangian conditions}
\label{sec:augLcond}

We give a more thorough discussion of the regularity conditions which ensure that the Augmented Lagrangian method will work. The goal of this section is simply to state these conditions and give intuitive arguments about why some should hold in our case, not to attempt to prove that they indeed hold.  The conditions \citep{bertsekas1996} are:

\begin{itemize}
\item There exists a strict local minimum $\phi^*$ of the optimization problem of Equation \ref{optobj}:

If the function class $\mathcal{F}$ is rich enough that it contains a true solver  of the maximum entropy problem, then a global optimum exists. Although not rigorous, we would expect that even in the finite expressivity case that a global optimum remains, and indeed, recent theoretical work \citep{raghu2016expressive,poole2016exponential} has gotten close to proving this.

\item $\phi^*$ is a regular point of the optimization problem, that is, the rows of $\nabla_\phi R(\phi^*)$ are linearly independent:

Again, this is not formal, but we should not expect this to cause any issues. This clearly depends on the specific form of $T$, but the condition basically says that there should not be redundant constraints at the optimum, so if $T$ is reasonable this shouldn't happen.

\item $H(p_\phi)$ and $R(\phi)$ are twice continuously differentiable on a neighborhood around $\phi^*$:

This holds by the smoothness of the normalizing flows.

\item $y^\top \nabla_\phi^2 L(\phi^*;\lambda^*,0) y>0$ for every $y\neq 0$ such that $\nabla_\phi R(\phi^*)y=0$, where $\lambda^*$ is the true Lagrange multiplier:

This condition is harder to justify. It would appear it is just asking that the Lagrangian (not the augmented Lagrangian) be strictly convex in feasible directions, but it is actually stronger than this and some simple functions might not satisfy the property. For example, if the function we are optimizing was $x^4$ and we had no constraints, the Lagrangian's Hessian would be $12x^2$, which is $0$ at $x^*=0$ thus not satisfying the condition. Importantly, these conditions are sufficient but not necessary, so even if this doesn't hold the augmented Lagrangian method might work (it certainly would for $x^4$). Because of this and the non-rigorous justifications of the first two conditions, we left these conditions for the appendix and relied instead on the empirical performance to justify that we are indeed recovering the maximum entropy distribution.

\end{itemize}

If all of these conditions hold, the augmented Lagrangian (for large enough $c$ and $\lambda$ close enough to $\lambda^*$) has a unique optimum in a neighborhood around $\phi^*$ that is close to $\phi^*$ (as $\lambda \rightarrow \lambda^*$ it converges to $\phi^*$) and its hessian at this optimum is positive-definite. Furthermore, $\lambda_k \rightarrow \lambda$. This implies that gradient descent (with the usual caveats of being started close enough to the solution and with the right steps) will correctly recover $\phi^*$ using the augmented Lagrangian method.  This of course just guarantees convergence to a local optimum, but if there are no additional assumptions such as convexity, it can be very hard to ensure that it is indeed a global optimum. Some recent research has attempted to explain why optimization algorithms perform so well for neural networks \citep{choromanska2015loss,kawaguchi2016deep}, but we leave such attempts for our case for future research.

\section{Risk-neutral asset price}
\label{sec:real}
We extract the risk-neutral asset price probability distribution based on option prices, an active and interesting area for ME models.  We give a brief introduction of the problem and refer interested readers to see \cite{buchen1996maximum} for a more detailed explanation. Denoting $S_t$ as the price of an asset at time $t$, the buyer of a European call option for the stock that expires at time $t_e$ with strike price $K$ will receive a payoff of $c_K = (S_{t_e} - K)_{+} = \max (S_{t_e} - K, 0)$ at time $t_e$. Under the efficient market assumption, the risk-neutral probability distribution for the stock price at time $t_e$ satisfies:
\begin{equation}
\label{equ:option_con1}
c_K = D(t_e) E_q[ (S_{t_e} - K)_{+} ],
\end{equation}
where $D(t_e)$ is the risk-free discount factor and $q$ is the risk-neutral measure. We also have that, under the risk-neutral measure, the current stock price $S_0$ is the discounted expected value of $S_{t_e}$:
\begin{equation}
\label{equ:option_con2}
S_0 = D(t_e) E_q(S_{t_e}).
\end{equation}
When given $m$ options that expire at time $t_e$ with strikes $K_1,...,K_m$ and prices $c_{K_1},...,c_{K_m}$, we get $m$ expectation constraints on $q(S_{t_e})$ from Equation \ref{equ:option_con1}, together with Equation \ref{equ:option_con2}, we have $m+1$ expectation constraints in total. With that partial knowledge we can approximate $q(S_{t_e})$, which is helpful for understanding the market expected volatility and identify mispricing in option markets, etc.

Inferring the risk-neutral density of asset price from a finite number of option prices is an important question in finance and has been studied extensively \citep{buchen1996maximum, borwein2003probability, bondarenko2003estimation, figlewski2008estimating}. One popular method proposed by \cite{buchen1996maximum} estimates the probability density as the maximum entropy distribution satisfying the expectation constraints and a positivity support constraint by fitting a Gibbs distribution, which results in a piece-wise linear log density:
\begin{equation}
p(z) \propto \exp \left\{ \eta_0 z + \sum_{i=1}^m \eta_i (z - K_i)_{+} \right\} \mathbbm{1}\left( z \geq 0 \right)
\end{equation}
and optimize the distribution with numerical methods. Here we compare the performance of the MEFN algorithm with the method proposed in \cite{buchen1996maximum}. To enforce the positivity constraint we choose $g(z)=e^{az+b}$, where $a$ and $b$ are additional parameters.

\begin{figure}[!ht]
\centering
\begin{tabular}[t]{ccc}
\vspace{-0cm}
\includegraphics[scale=0.43,clip = true]{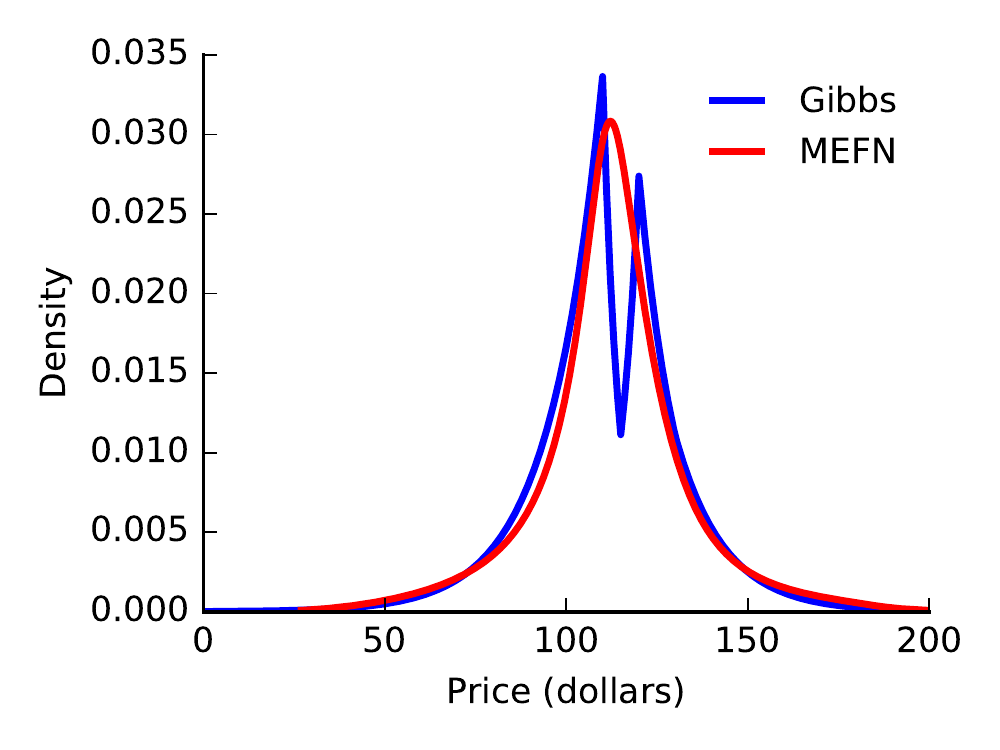}&
\includegraphics[scale=0.43,clip = true]{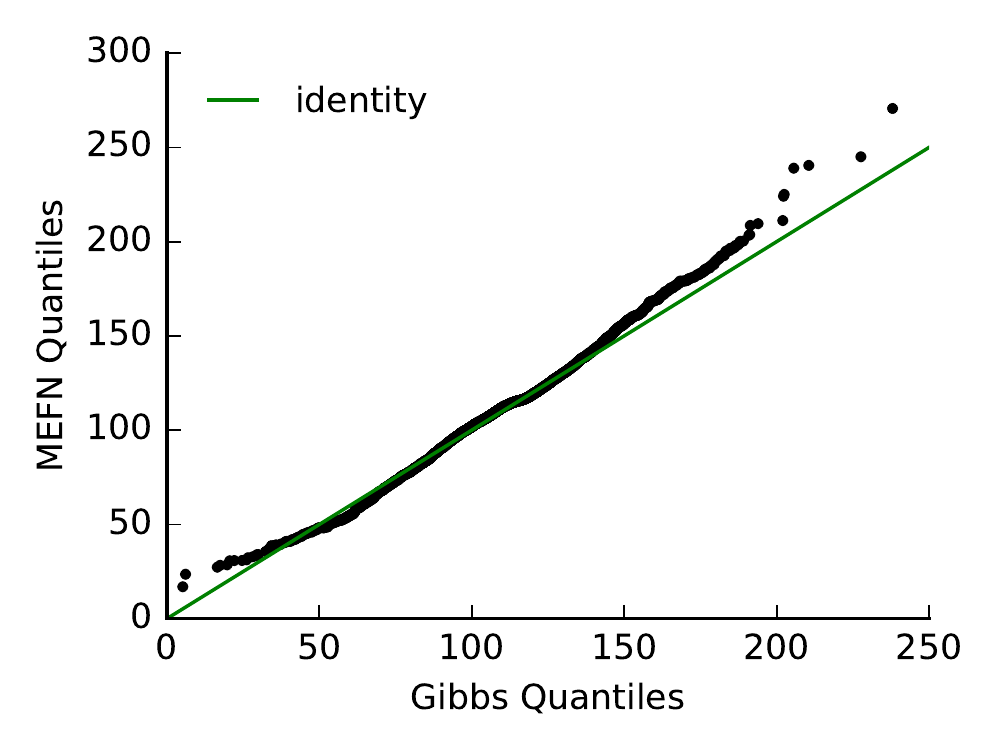}&
\includegraphics[scale=0.43,clip = true, trim=0cm 0cm 0cm 0cm]{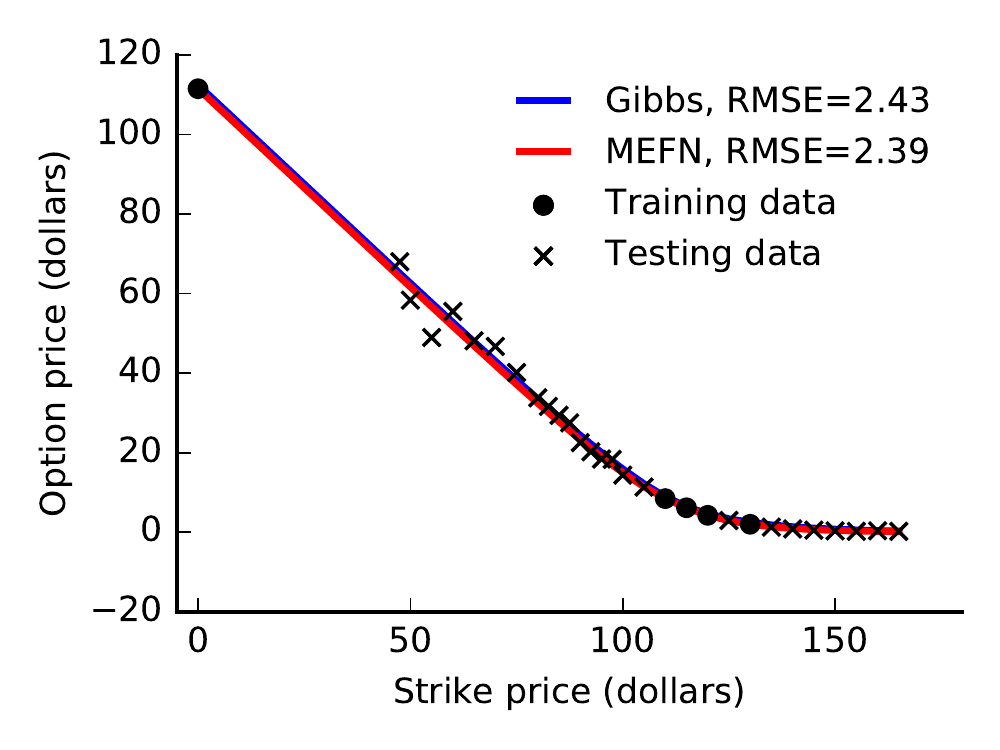}\\
%\\
\end{tabular}
\caption{Constructing risk-neutral measure from observed option price. \textit{Left panel}: fitted risk-neutral measure by Gibbs and MEFN method. \textit{Middle panel}: Q-Q plot for the quantiles from the distributions on the left panel. \textit{Right panel}: observed and fitted option price for different strikes.}
\label{fig:option}
%\label{fig:reaching_projection}
\end{figure}

We collect the closing price of European call options on Nov. 1 2016 for the stock AAPL (Apple inc.) that expires on $t_e = $ Jun. 16 2017. We use $m=4$ of the options with highest trading volume as training data and the rest as testing data. On the left panel of figure \ref{fig:option}, we show the fitted risk-neutral density of $S_{t_e}$ by MEFN (red line) with that of the fitted Gibbs distribution result (blue line). We find that while the distributions share similar location and variability, the distribution inferred by MEFN is smoother and arguably more plausible. In the middle panel we show a Q-Q plot of the quantiles of the MEFN and Gibbs distributions. We can see that the quantile pairs match the identity closely, which should happen if both methods recovered the exact same distribution. This highlights the effectiveness of MEFN.  There does exist a small mismatch in the tails: the distribution inferred by MEFN has slightly heavier tails. This mismatch is difficult to interpret: given that both the Gibbs and MEFN distributions are fit with option price data (and given that one can observe at most one value from the distribution, namely the stock price at expiration), it is fundamentally unclear which distribution is superior, in the sense of better capturing the true ME distribution's tails. On the right panel we show the fitted option price for the two fitted distributions (for each strike price, we can recover the fitted option price by Equation \ref{equ:option_con1}). We noted that the fitted option price and strike price lines for both methods are very similar (they are mostly indiscernible on the right panel of figure \ref{fig:option}). We also compare the fitted performance on the test data by computing the root mean square error for the fitted and test data. We observe that the predictive performances for both methods are comparable. 

We note that for this specific application, there are practical concerns such as the microstructure noise in the data and inefficiency in the market, etc. Applying a pre-processing procedure and incorporating prior assumptions can be helpful for getting a more full-fledged method (see e.g. \cite{figlewski2008estimating}). Here we mainly focus on illustrating the ability of the MEFN method to approximate the ME distribution for non-typical distributions. Future work for this application includes fitting a risk-neutral distribution for multi-dimensional assets by incorporating dependence structure on assets.

\section{Modeling images of textures}
\label{sec:textapp}

\begin{figure}[!ht]
\centering
\hspace*{-0.4cm}
\begin{tabular}{cccccc}
\multicolumn{2}{c}{\hspace{-0.5cm}\begin{tabular}[x]{@{}c@{}}\vspace{-0.1cm} Input\\  (positive example)\end{tabular}} & \multicolumn{2}{c}{\hspace{-0.5cm}\begin{tabular}[x]{@{}c@{}}\vspace{-0.1cm}Input\\ (positive example)\end{tabular}} & \multicolumn{2}{c}{\hspace{-0.5cm}\begin{tabular}[x]{@{}c@{}}\vspace{-0.1cm}Input\\ (negative example)\end{tabular}} \\
\multicolumn{2}{c}{\includegraphics[scale=0.18,clip = true, trim={5cm 2cm 2cm 1cm}]{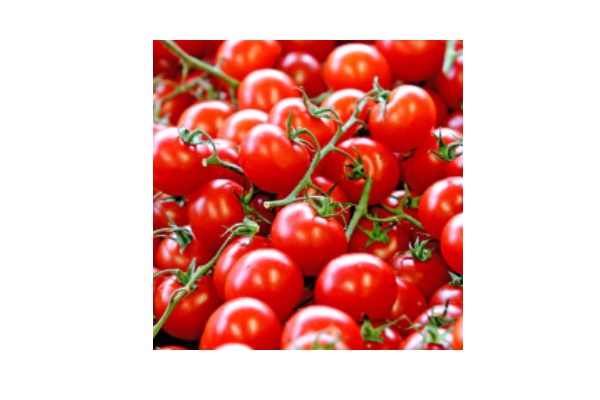}} & \multicolumn{2}{c}{\includegraphics[scale=0.18,clip = true, trim={5cm 2cm 2cm 1cm}]{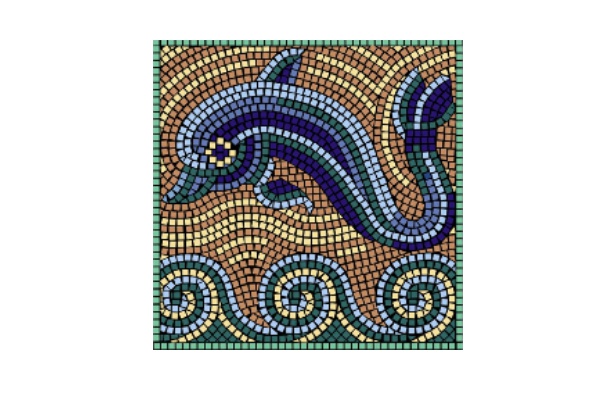}} & \multicolumn{2}{c}{\includegraphics[scale=0.18,clip = true, trim={5cm 2cm 2cm 1cm}]{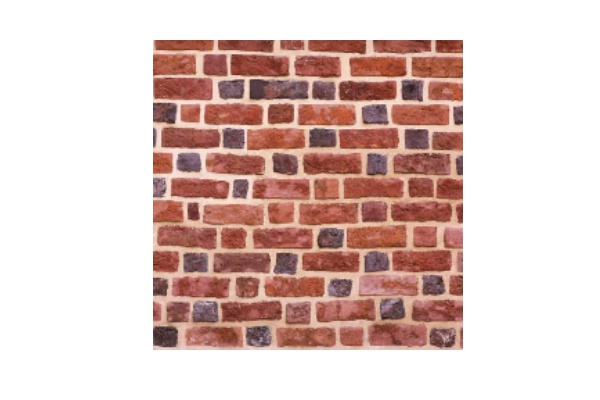}} \\
 
\multicolumn{6}{c}{Texture net (\cite{ulyanov2016texture}, less sample diversity)}\\
\includegraphics[scale=0.18,clip = true, trim={5cm 2cm 2cm 1cm}]{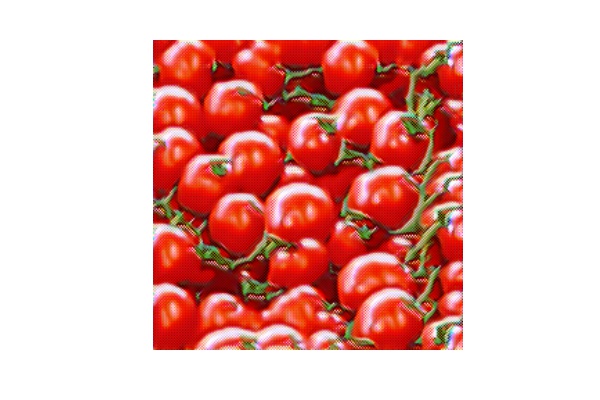}\hspace{-0.4cm} & \hspace{-0.4cm}\includegraphics[scale=0.18,clip = true, trim={5cm 2cm 2cm 1cm}]{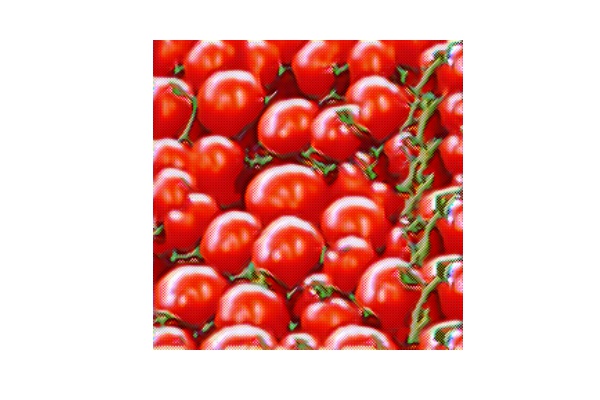} & \includegraphics[scale=0.18,clip = true, trim={5cm 2cm 2cm 1cm}]{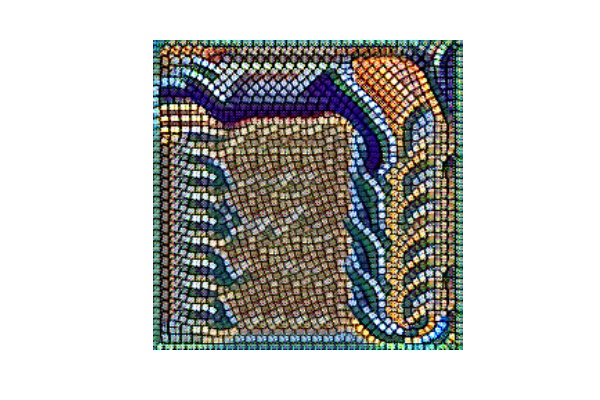}\hspace{-0.4cm} & \hspace{-0.4cm}\includegraphics[scale=0.18,clip = true, trim={5cm 2cm 2cm 1cm}]{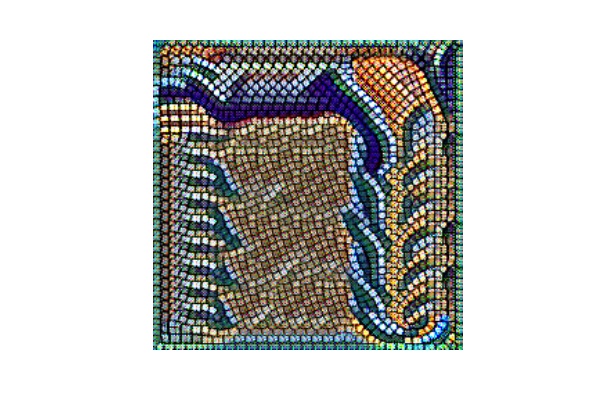} & \includegraphics[scale=0.18,clip = true, trim={5cm 2cm 2cm 1cm}]{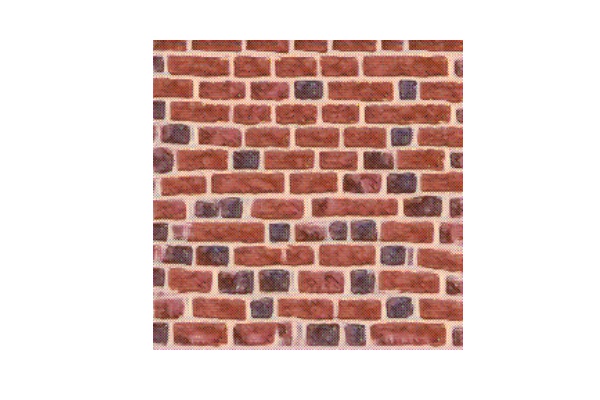}\hspace{-0.4cm} & 
\hspace{-0.4cm}\includegraphics[scale=0.18,clip = true, trim={5cm 2cm 2cm 1cm}]{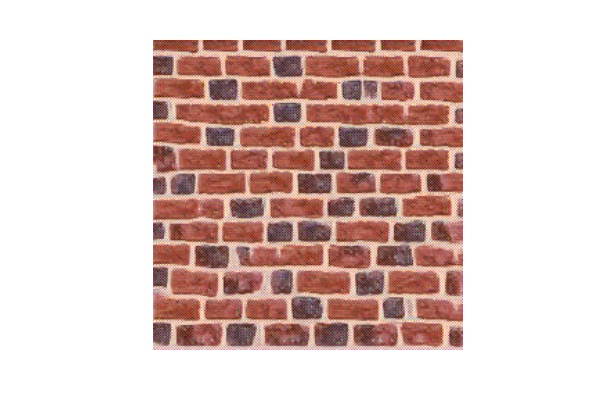}\\
\includegraphics[scale=0.18,clip = true, trim={5cm 2cm 2cm 1cm}]{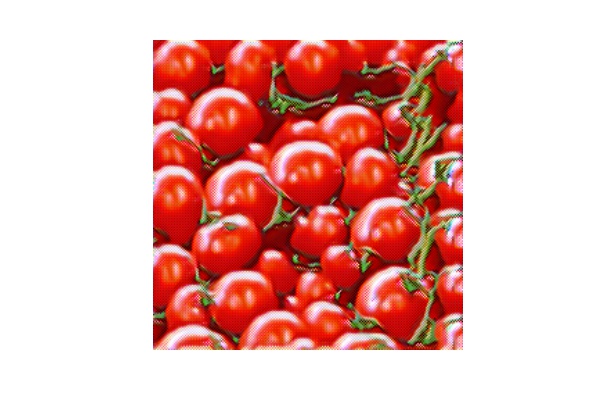}\hspace{-0.4cm} & \hspace{-0.4cm}\includegraphics[scale=0.18,clip = true, trim={5cm 2cm 2cm 1cm}]{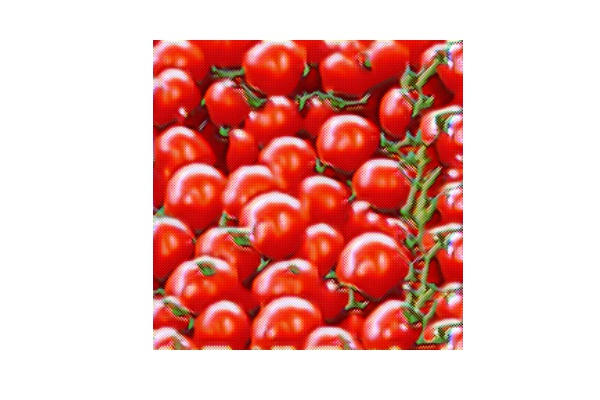} & \includegraphics[scale=0.18,clip = true, trim={5cm 2cm 2cm 1cm}]{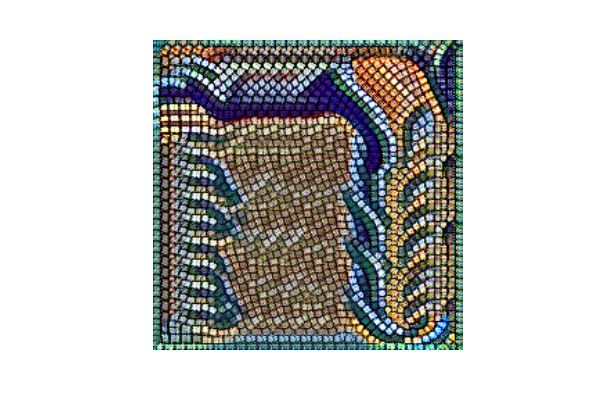}\hspace{-0.4cm} & \hspace{-0.4cm}\includegraphics[scale=0.18,clip = true, trim={5cm 2cm 2cm 1cm}]{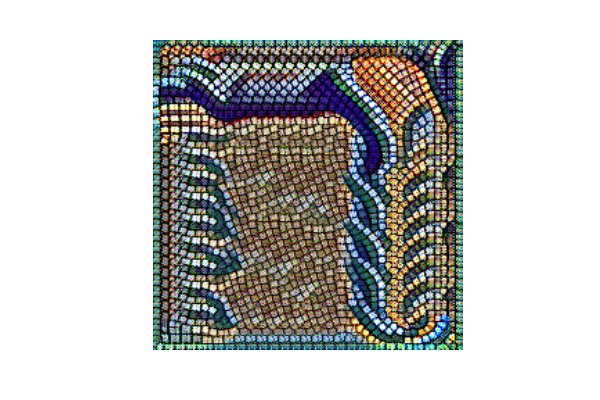} & \includegraphics[scale=0.18,clip = true, trim={5cm 2cm 2cm 1cm}]{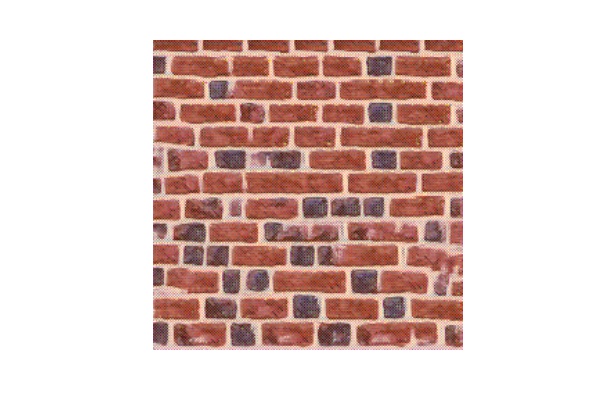}\hspace{-0.4cm} & 
\hspace{-0.4cm}\includegraphics[scale=0.18,clip = true, trim={5cm 2cm 2cm 1cm}]{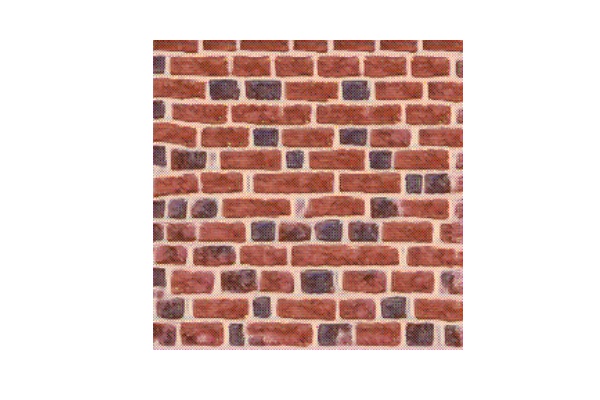}\\

\multicolumn{6}{c}{MEFN (ours, more sample diversity)}\\

\includegraphics[scale=0.18,clip = true, trim={5cm 2cm 2cm 1cm}]{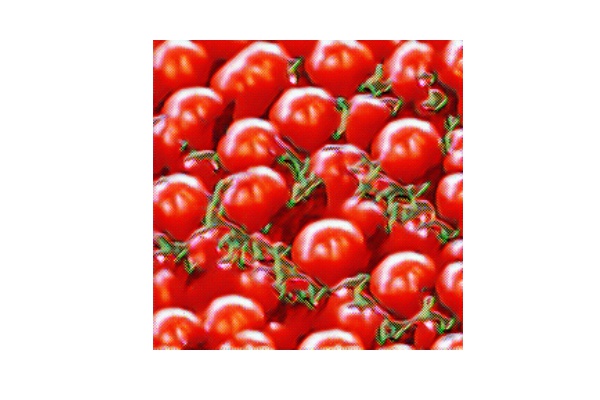}\hspace{-0.4cm} & \hspace{-0.4cm}\includegraphics[scale=0.18,clip = true, trim={5cm 2cm 2cm 1cm}]{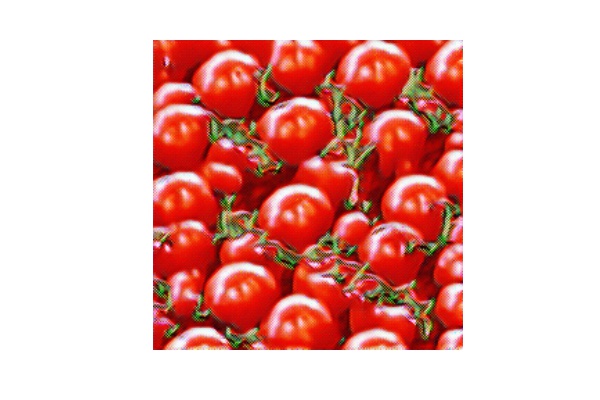} & \includegraphics[scale=0.18,clip = true, trim={5cm 2cm 2cm 1cm}]{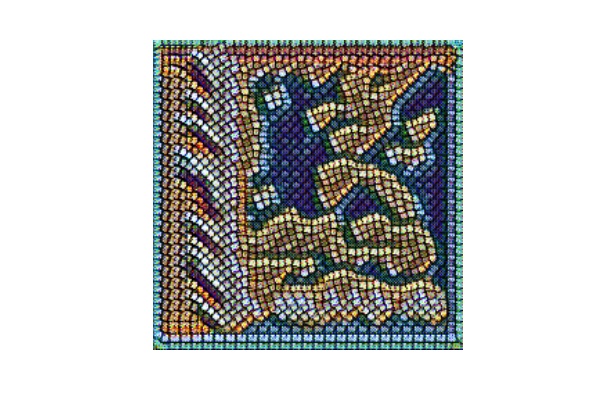}\hspace{-0.4cm} & \hspace{-0.4cm}\includegraphics[scale=0.18,clip = true, trim={5cm 2cm 2cm 1cm}]{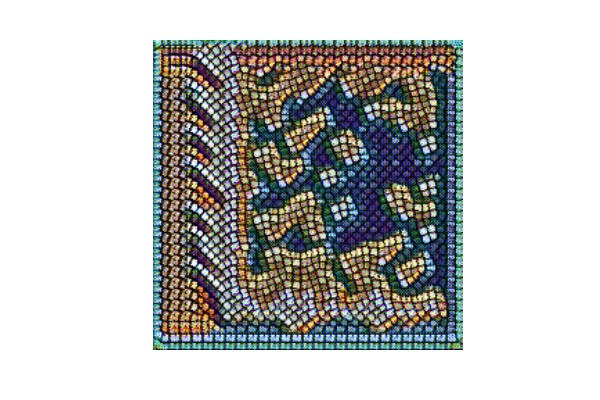} & \includegraphics[scale=0.18,clip = true, trim={5cm 2cm 2cm 1cm}]{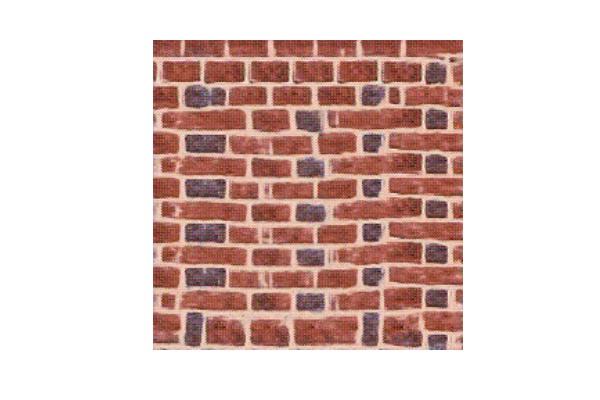}\hspace{-0.4cm} & 
\hspace{-0.4cm}\includegraphics[scale=0.18,clip = true, trim={5cm 2cm 2cm 1cm}]{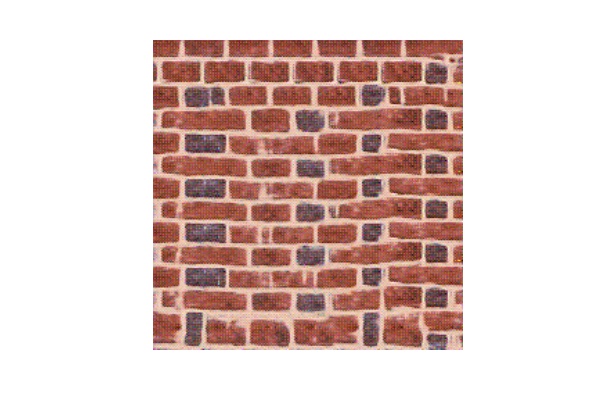}\\
\includegraphics[scale=0.18,clip = true, trim={5cm 2cm 2cm 1cm}]{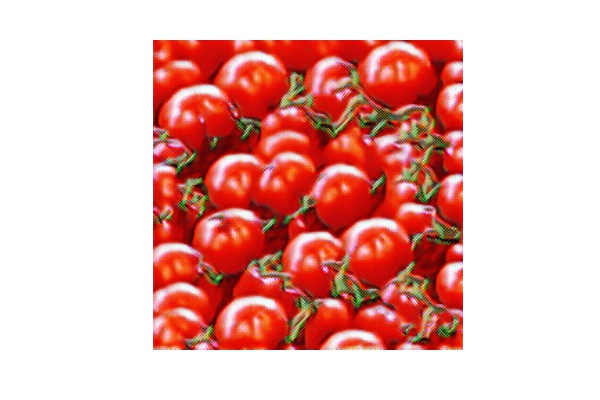}\hspace{-0.4cm} & \hspace{-0.4cm}\includegraphics[scale=0.18,clip = true, trim={5cm 2cm 2cm 1cm}]{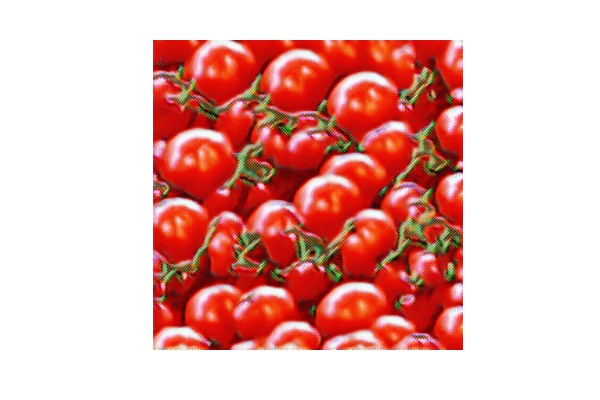} & \includegraphics[scale=0.18,clip = true, trim={5cm 2cm 2cm 1cm}]{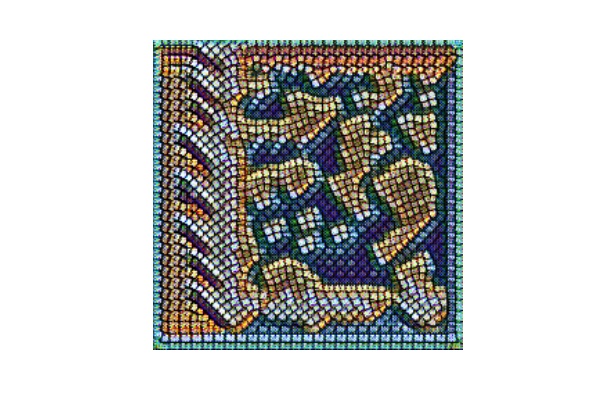}\hspace{-0.4cm} & \hspace{-0.4cm}\includegraphics[scale=0.18,clip = true, trim={5cm 2cm 2cm 1cm}]{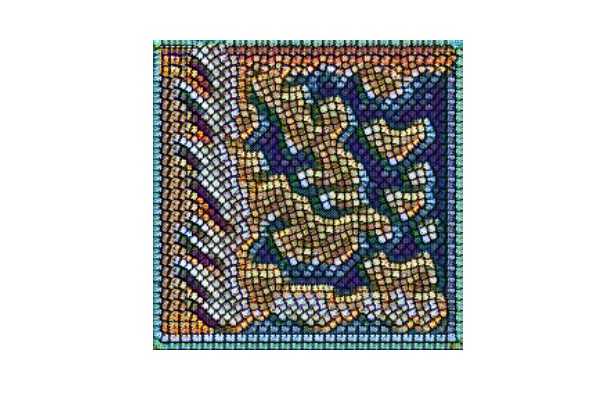} & \includegraphics[scale=0.18,clip = true, trim={5cm 2cm 2cm 1cm}]{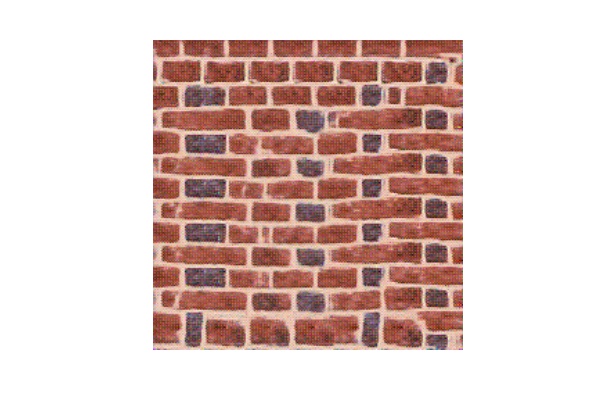}\hspace{-0.4cm} & 
\hspace{-0.4cm}\includegraphics[scale=0.18,clip = true, trim={5cm 2cm 2cm 1cm}]{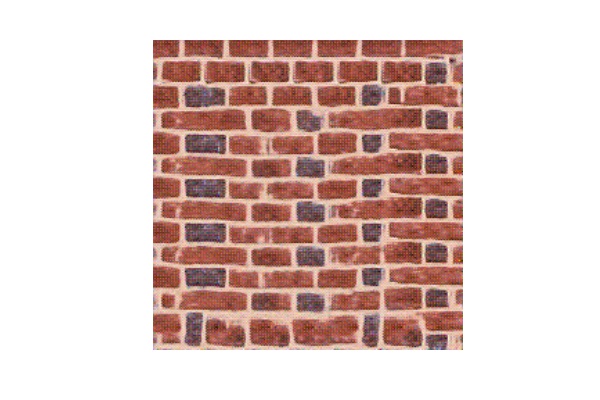}

\end{tabular}
\caption{MEFN and texture network samples.}
\label{fig:textapp}
\end{figure}

We tried our texture modeling approach with many different textures, and although MEFN samples don't always exhibit more visual diversity than samples obtained from the texture network, they always have more entropy as in figure \ref{fig:texture}.  Figure \ref{fig:textapp} shows two positive examples, i.e. textures in which samples from MEFN do exhibit higher visual diversity than those from the texture network, as well as a negative example, in which MEFN achieves less visual diversity than the texture network, regardless of the fact that MEFN samples do have larger entropy. We hypothesize that this curious behavior is due to the optimization achieving a local optimum in which the brick boundaries and dark brick locations are not diverse but the entropy within each brick is large. It should also be noted that among the experiments that we ran, this was the only negative example that we got, and that slightly modifying the hyperparameters caused the issue to disappear. %\footnote{Appendix section \S \ref{sec:textapp} does not appear on the ICLR version of the paper since the results were obtained after the camera-ready deadline.}

\end{document}